\documentclass[useAMS,usenatbib,usegraphicx]{mn2e}

\usepackage{epsf,amsmath,rotating}

\def\apj{ApJ}
\def\apjs{ApJS}
\def\aj{AJ}
\def\mnras{MNRAS}
\def\aa{A\&A}
\def\an{AN}
\newcommand{\bq}{\begin{equation}}
\newcommand{\eq}{\end{equation}}
\newcommand{\bqn}{\begin{eqnarray}}
\newcommand{\eqn}{\end{eqnarray}}
\newcommand{\dd}{\mbox{\rm d}}
\newcommand{\msun}{\rm{M}_\mathrm{\rm \sun}}

\title[Milky Way disc model - II]
{Towards a fully consistent Milky Way disc model - II.
The local disc model and SDSS data of the NGP region}

\author[A. Just et al.]{A. Just$^1$, S. Gao$^1$, S. Vidrih$^{1,2}$\\
  $^{1}$ Astrononisches Rechen-Institut, Zentrum f\"ur Astronomie der
    Universit\"at Heidelberg (ZAH), 
  M\"onchhofstr. 12-14, 69120 Heidelberg, Germany \\
  $^{2}$ University of Ljubljana, Faculty of Mathematics and Physics, 
  Department of Physics, Jadranska 19, 1000 Ljubljana, Slovenia}

\begin{document} 

\maketitle

\begin{abstract}
We have used the self-consistent vertical disc models of the solar neighbourhood  presented in \citet{Ju10},
which are based on different star formation histories (SFR) and fit the local kinematics of main sequence stars equally well, to predict star counts towards the North Galactic Pole (NGP). We
combined these four different models with
the local main sequence in the filter system of the SDSS and predicted the star
counts in the NGP field with $b>80\deg$. All models fit the Hess diagrams in the
F--K dwarf regime better than $\pm 20$ percent and the star number densities in
the solar neighbourhood are consistent with the observed values.
The $\chi^2$ analysis shows that model A is clearly preferred with systematic
deviations of a few percent only. The SFR of model A is characterised by a
maximum at an age of 10\,Gyr and a decline by a factor of four to the present day
value of 1.4\,$\msun$/pc$^2$/Gyr. 
The thick disc can be modelled very well by an old isothermal simple stellar
population. The density profile can be approximated by a
sech$^{\alpha_\mathrm{t}}$ function. We found a power law index 
$\alpha_\mathrm{t}=1.16$ and a scale height
$h_\mathrm{t}=800$\,pc corresponding to a vertical velocity dispersion of 
$\sigma_\mathrm{t}=45.3$\,km/s. About 6 percent of the stars in the solar
neighbourhood are thick disc stars. 
\end{abstract}

\begin{keywords}
Galaxy: solar neighbourhood -- 
Galaxy: disc -- Galaxy: structure -- Galaxy: evolution -- Galaxy: stellar 
content -- Galaxy: kinematics and dynamics
\end{keywords}

\section{Introduction\label{sec-intro}}

In \citet{Ju10} (hereafter Paper I) a self-consistent model of the vertical structure of the Milky
Way disc in the solar neighbourhood was presented. The model is based on the
star formation history (SFR), the age velocity dispersion relation (AVR), and a
simple chemical enrichment model. The vertical density profiles of the stellar
sub-populations are self-consistently calculated in the total gravitational
potential of stars including the contribution of the gas component and the dark
matter halo. The input parameters are selected and optimised to reproduce the
velocity distribution functions $f_\mathrm{i}(W)$ for the vertical velocity component $W$ along the main sequence
(MS). It turned out that for each SFR the range of fitting AVRs is very small.
On the other hand the SFR is not well determined by the local kinematics only.
Therefore we discussed in Paper I four different SFRs with similar $\chi^2$
values to cover the range of possible functional shapes (models A, B, C, and D).
The overall range of best fitting AVRs is still well restricted. Due to the
different age distributions and consequently different relative contributions of
the sub-populations as function of scale height the vertical density profiles
of MS stars differ significantly at large distances $z$ above the Galactic mid-plane.

A combination of the local model with number densities of MS stars in the solar neighbourhood
allows predictions of star counts at high Galactic latitude. Therefore we can
use star counts of large surveys for an independent test
of the local model. As a new additional result we expect to find restrictions on
the SFR. As local normalisation along the MS we use Hipparcos stars
complemented by the Catalogue of Nearby Stars (CNS4) at the faint end \citep{jah97}. 
The model predictions will be compared to the North Galactic Pole (NGP) data 
of the Sloan Digital Sky Survey (SDSS), 
which has collected at present the largest and most homogeneous database 
comprising about $10^8$ stellar objects in the Milky Way \citep{Gu98,Ab09}. 
The SDSS photometry is in the $u\, g\, r\, i\, z$ filter system.
Despite the enormous wealth of information that the SDSS database provides 
in the $u\,g\,r\,i\,z$ filter system, the
majority of current observational and theoretical knowledge of resolved stellar
populations is based largely on the Johnson-Kron-Cousins $UBVR_CI_C$ photometric
system and some other systems such as the Str\"omgren, DDO, Vilnius, and Geneva
systems. To overcome this difficulty, we use an
empirical transformation of the nearby MS star photometry into the 
$u\,g\,r\,i\,z$ system \citep{Ju08}.
For each theoretical model we derive a best fit solution for the full Hess diagrams in $(g-r,g)$, which quantify the number density distributions in the colour
magnitude diagram (CMD).

The paper is organised as follows.
In Sect.\ \ref{sec-data} we describe the data selection, 
in Sect.\ \ref{sec-model} important properties of the theoretical models are presented, 
in Sect.\ \ref{sec-local} the local normalisation is discussed,
in Sect.\ \ref{sec-fit} we present the best fit procedure,
in Sect.\ \ref{sec-results} we discuss the
results,
and in Sect.\ \ref{sec-conclusion} we draw some conclusions.

\section{SDSS Data}\label{sec-data}

The Sloan Digital Sky Survey (SDSS) is an imaging and spectroscopic 
survey~\citep{Yo00} that has since 1998 with its $2.5\,$m dedicated telescope 
mapped more than a quarter of the sky \citep{Gu98}. 
Photometric sky coverage of the SDSS Data
Release 7 (DR7) amounts to $11,663\,\mathrm{deg^2}$, including a
$7,646\,\mathrm{deg^2}$ large contiguous area around the NGP~\citep{Am08,Ab09}. 

\citet{Sm02} defined the $u'\, g'\, r'\, i'\, z'$ photometric system on 158 
standard stars, a subset of $UBVR_CI_C$ standard stars from~\citet{La92}, using
the USNO-1.0m telescope. Unfortunately, the photometric system of the 2.5m SDSS 
telescope, denoted as $u\, g\, r\, i\, z$, slightly differs from the
$u'\, g'\, r'\, i'\, z'$ one~\citep{Ab03}. Moreover, the SDSS standards are too
bright and saturate in the 2.5m SDSS telescope during its normal operational
mode. These inconveniences have been resolved by using fainter secondary 
standards scattered throughout the SDSS survey area and by using simple linear
transformation equations between the two photometric sets~\citep{Tu06}. The 
nightly photometry obtained by the SDSS 2.5m telescope can be thus calibrated to
the native $u\, g\, r\, i\, z$ system with magnitude zero points accurate on the
AB system to within a few percent.

Imaging data are produced simultaneously in the five photometric 
bands, namely $u$, $g$, $r$, $i$, and $z$~\citep{Fu96,Gu06,Sm02,Ho01}. The 
images are automatically processed through specialised 
pipelines~\citep{Lu99,Lu01,St02,Pi03,Tu06} producing corrected images, object 
catalogues, astrometric solutions, calibrated fluxes and many other data 
products. SDSS photometry is homogeneous and deep ($r<22.5$), repeatable to 
$0.02\,$mag~\citep{Iv03} and with a zero-point uncertainty of 
$\sim 0.01-0.02$~\citep{Ab04,Iv04}. In DR7 a homogeneous photometry over the full sky coverage was established at the 1 percent level in $g r i z$ and 2 percent in $u$ in a process called ubercalibration \citep{Pa08}.

We have limited our analysis to the stars in the magnitude range $14\leq g
\leq 20.5$\footnote{Magnitude limits applied to the dereddenend magnitudes. 
We applied the standard dereddening of the data based on the extinction map of
\citet{Sc98} as given in the data base of SDSS.}. 
For brighter stars the CCD camera of the SDSS telescope saturates.
At the faint end we wanted to completely avoid the problems in the galaxy-star
separation and we set a conservative magnitude limit for this purpose. The SDSS 
photometric data contain also quality flags for each object to aid in the 
selection of "good" measurements\footnote{See the following link for the clean 
photometry prescriptions:\\
http://cas.sdss.org/dr7/en/help/docs/realquery.asp\#flags.}. We have
carefully analysed the appearance of these photometric flags with respect to the
object brightness. We concluded that the problematic flags relate mostly to the 
stars fainter than our faint magnitude limit. Many of these flags are also 
tightly correlated with the magnitude measurement error. To make our stellar 
samples as complete as possible we have finally applied only one "cleaning" 
criterion, i.e. we have rejected all the measurements with reported magnitude 
error larger than $0.2$ in $g$ or in $r$ filter, which typically accounted 
for 1 percent or even less of the total star counts. It is also necessary to 
mention that the typical magnitude error in the chosen magnitude range is much 
smaller than the applied magnitude error limit.

We selected the colour range $-0.2 < g-r < 1.2$ with a symmetric bin size of
$\pm 0.025$\,mag. This range covers the MS down to K dwarfs, where the local
normalisation is reliable.
In order to test the predictions from the vertical structure of the local disc 
model (Paper I) with the available SDSS
photometric data we have selected a field at the NGP. The field should not reach
too low galactic latitude in order to avoid projection effects and a dependence
of the star counts on the radial properties of the disc model.
At the same time the field should be as large as possible in order to reduce the 
Poisson noise of the star counts. We have chosen a NGP field with
Galactic latitudes $b>80^\circ$ and an area of 
$A_\mathrm{NGP}=313.36\,\mathrm{deg}^2$. In $g$ magnitude
we use a resolution of $\dd g=0.01$\,mag, but for the fitting procedure we applied a
car-box smoothing of $\Delta g=$0.1, 0.2, and 0.5\,mag, where the last value
is our standard case. The resolution in colour is $\dd (g-r)=0.05$\,mag. The Hess diagram of the data is
shown in the bottom panel of figure \ref{figcomp}.

In order to avoid confusion of observational errors in the SEGUE data
and uncertainties of the model predictions we compare dereddened number counts
$Y(g-r,g)$ in the colour -- apparent magnitude plane. 

\section{The Milky Way Model}\label{sec-model}

We use the self-consistent local model of the thin and thick disc described in 
detail in Paper I and add a simple stellar halo component.
In this section we report the necessary features to understand the
procedure applied here.
The disc model relies on the kinematics of MS stars in the solar
neighbourhood measured by volume complete local samples of stars combining 
Hipparcos stars and faint stars of the CNS4.
The properties of the model depend on the total stellar mass density in the solar
neighbourhood and only weakly on the adopted Initial Mass Function (IMF).
It is independent of star counts at large distances.

\begin{figure}
\centerline{\resizebox{0.98\hsize}{!}{\includegraphics[angle=0]{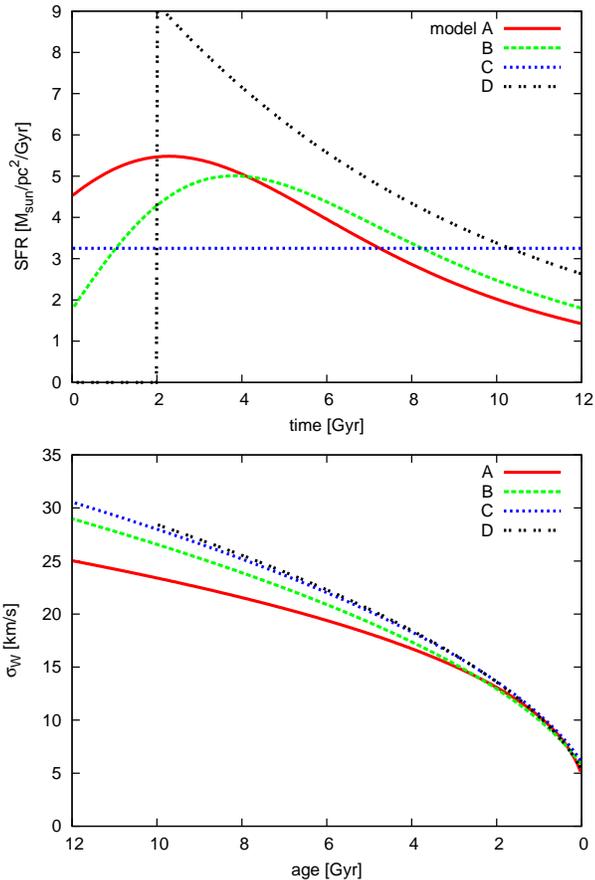}}}
\caption[]{
The upper panel shows the SFRs as function of time and
the lower panel gives the AVRs of models A -- D as function of age (age is running backwards).
}
\label{figsfr}
\end{figure}

\begin{figure}
\centerline{\resizebox{0.98\hsize}{!}{\includegraphics[angle=270]{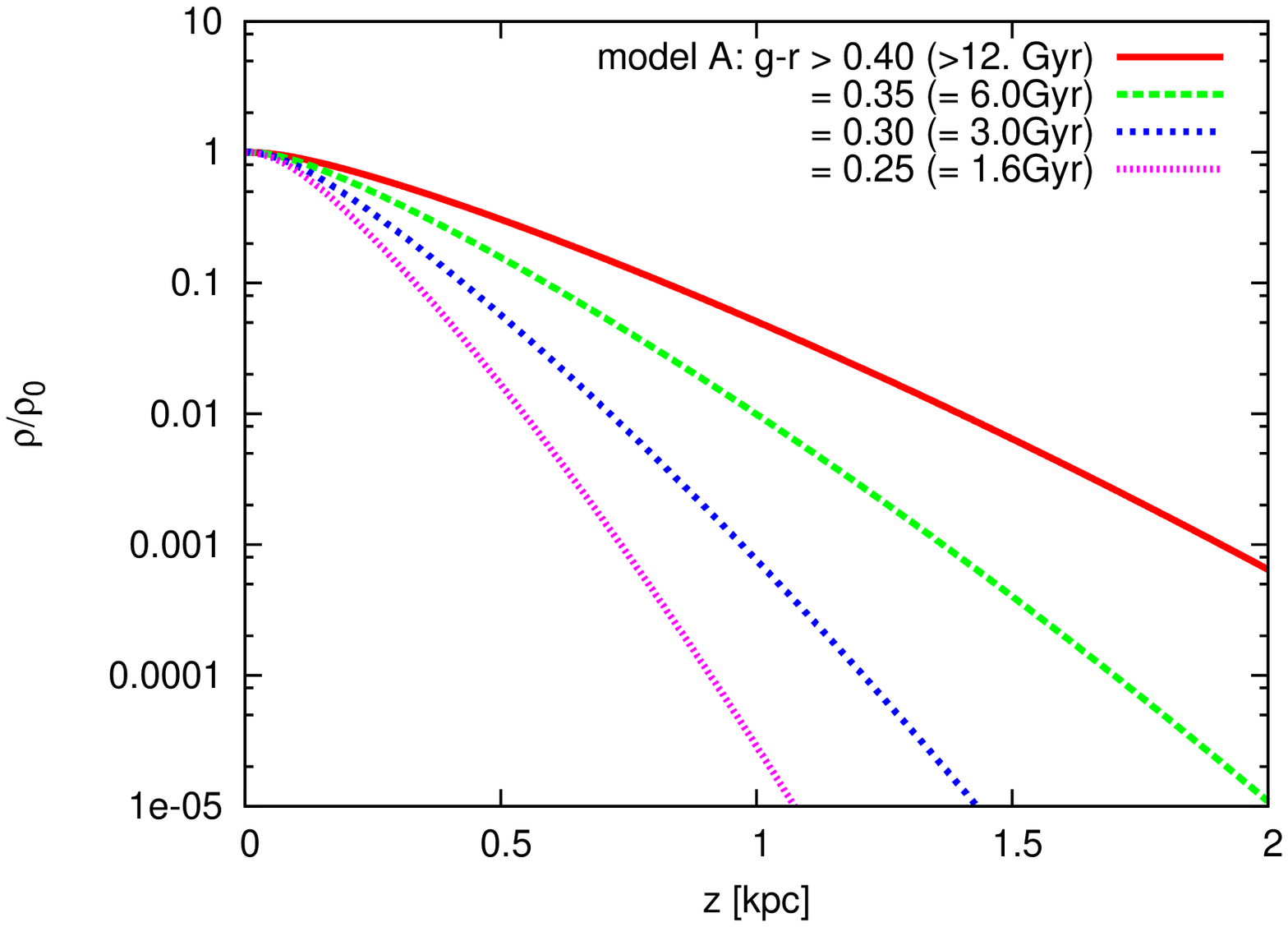}}}
\centerline{\resizebox{0.98\hsize}{!}{\includegraphics[angle=270]{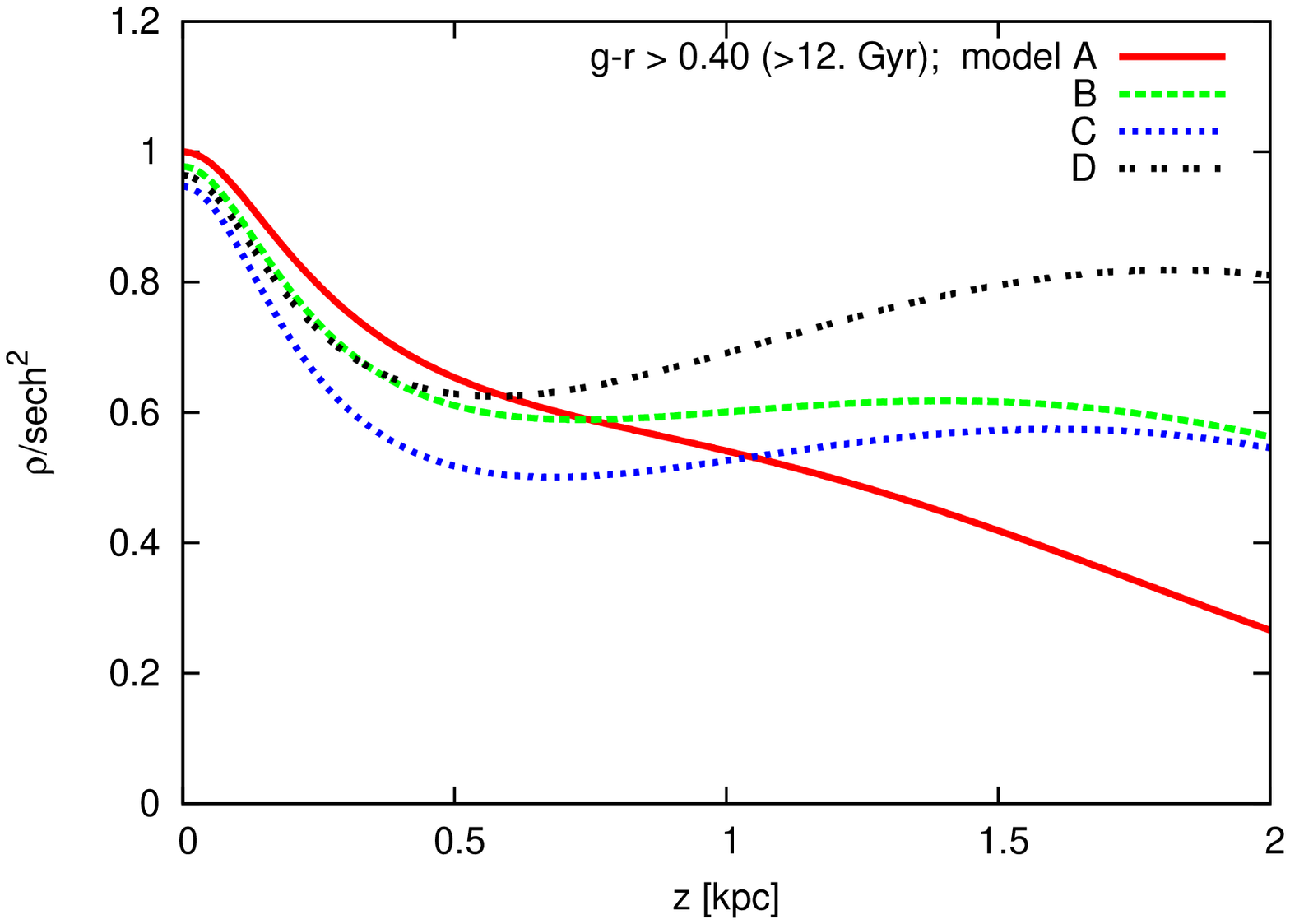}}}
\centerline{\resizebox{0.98\hsize}{!}{\includegraphics[angle=270]{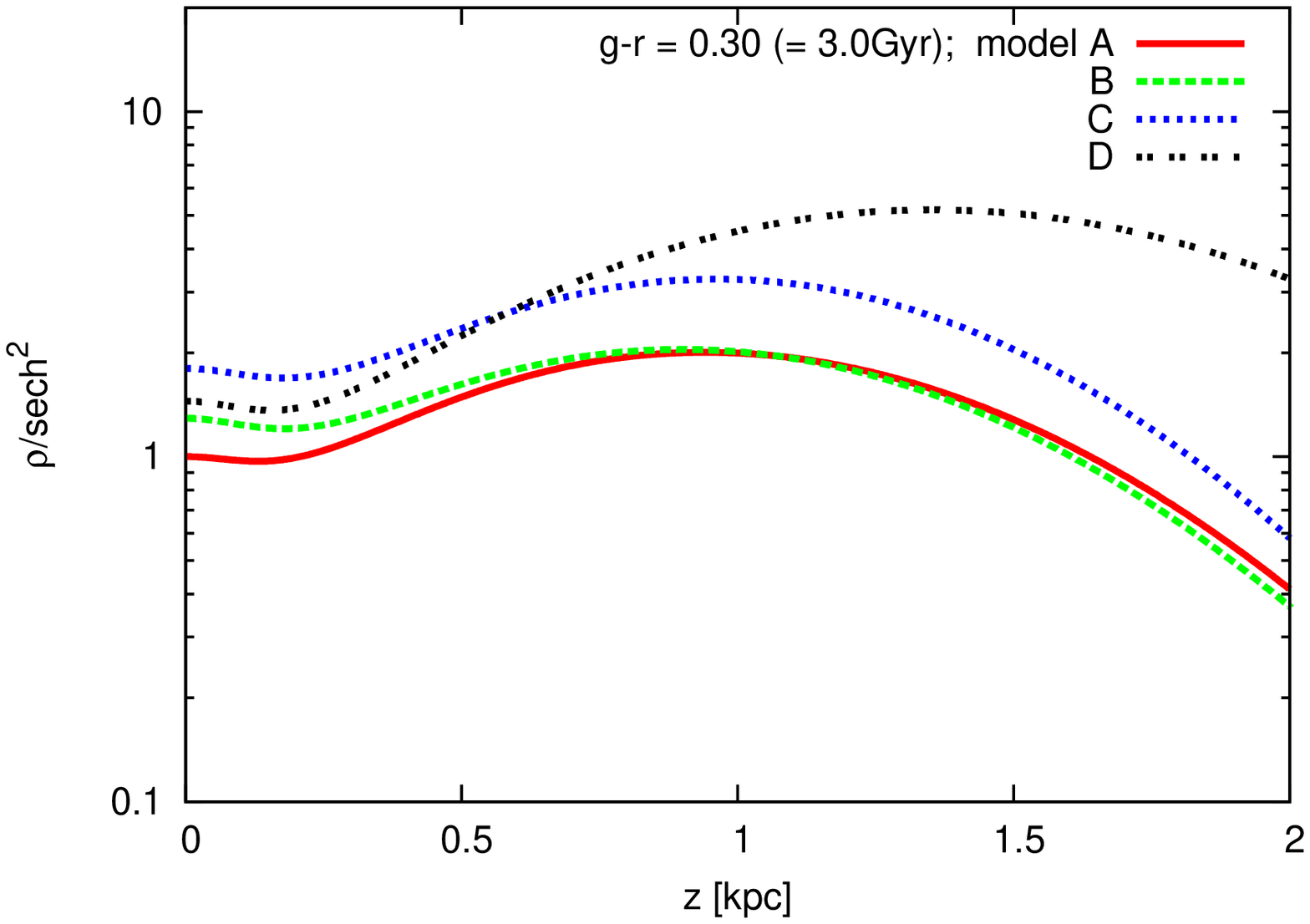}}}
\caption[]{
The top panel shows normalised density profiles of model A for different colour bins
(with the corresponding MS lifetime in parenthesis). 
The middle panel shows the deviations of the density profiles for models A -- D from a
$\rho_0\mbox{sech}^2(z/2\,z_\mathrm{s})$ profile with $z_\mathrm{s}=270$\,pc for 
stars with lifetime larger than 12\,Gyr ($g-r>0.4$).
The lower panel is in log-scale and shows the same for 
a lifetime of 3\,Gyr using $z_\mathrm{s}=108$\,pc.
}
\label{figrhoms}
\end{figure}

The thin disc model at the solar radius is based on a pair
(SFR,AVR) as input functions, which determines the age distribution and the velocity distribution 
functions of the stellar sub-populations.
The vertical density profiles are calculated self-consistently 
in dynamical equilibrium in the
total gravitational potential of stars, gas and dark matter halo.
The chemical enrichment is also included for the determination of MS
lifetimes, colour indices and luminosities from Padova population synthesis
models.
For each pair of functions (SFR,AVR) the local disc model provides a unique 
connection of local star counts
and the number density of stars above the galactic plane with no additional free
parameters. In Paper I we compared four different models A, B, C, D which all
yield similar best fit $\chi^2$ values. The SFR and AVR of these models are
plotted in figure \ref{figsfr}.
The density profile of MS stars at a given colour index is characterised by the MS
lifetime of the stars. It is composed of a series of
density profiles of coeval stars according to the SFR up to the lifetime. 
The top panel of figure \ref{figrhoms} shows the
normalised density profiles of model A for the relevant colour index range in $g-r$. Bluer thin disc stars are too bright to be visible in SDSS at high Galactic latitude. 

The lower panels of figure \ref{figrhoms} show the density profiles of models A -- D
 divided by a
$\mbox{sech}^2(z/2\,z_\mathrm{s})$ profile with exponential scale height 
$z_\mathrm{s}$ for two colour (MS lifetime) bins. For both lifetimes
we use the corresponding exponential scale height of model A, i.e.
$z_\mathrm{s}$=270, 108\,pc respectively. The middle panel is in linear
scale, whereas the lower panel for stars with shorter lifetime is in log-scale.
The deviations from a sech$^2$-profile of model A are significant for all lifetimes,
because the populations are not isothermal and their spatial distribution is influenced by the gravitational
potential of all other components. For younger populations the deviations
are much stronger than for older populations.
Between the different models the differences exceed a factor of two. Models B--D
with a lower fraction of stars older than 10\,Gyr require larger velocity
dispersions of the old thin disc populations (see figure \ref{figsfr}). This leads
to shallower density profiles at $z>1$\,kpc.
All profiles are based on a Scalo IMF. The differences at the mid-plane $z=0$ can be
corrected by adjusting the IMF, which will be done implicitly by fitting the local
normalisation (see below). In any case large differences of star counts as function of
distance (apparent magnitude) remain.

In each $g-r$ colour bin the density profile is characterised by the MS
lifetime. According to Paper I and \citet{Ju08} we use for the bins $g-r$=0.15, 0.2,
0.25 the lifetime 1.2, 1.4, 1.6\,Gyr, respectively. In the colour bins
$g-r$=0.3, 0.35, 0.4 we include the contribution of turnoff stars and use
lifetime ranges 2.4--3.2, 4.0--6.2 and 8.2--12\,Gyr, respectively. For all colours
$g-r\ge$0.45 the lifetime is larger than 12\,Gyr leading to the same density
profile.

An isothermal thick disc component is
self-consistently included in the local model. We have shown in Paper I that it
can be well fitted by a
\bq
\rho_\mathrm{t}(z)=\rho_\mathrm{t,0}\mathrm{sech}^{\alpha_\mathrm{t}}
	[z/(\alpha_\mathrm{t}\,z_\mathrm{t})] 
\label{eq-rhot}
\eq
law with local density 
$\rho_\mathrm{t,0}$ and exponential scale height $z_\mathrm{t}$. Since the influence of
the thick disc on the total gravitational potential $\Phi(z)$ is very small, 
the thick disc parameters can be varied in a large range with negligible influence on 
$\Phi(z)$ and the thin disc structure. From the Jeans equation it follows that 
$z_\mathrm{t}$ and $\alpha_\mathrm{t}^{-1}$ are both proportional to the 
square of the velocity dispersion $\sigma_\mathrm{t}$ of
the thick disc in a given gravitational potential. As a consequence we find
\bq
\alpha_\mathrm{t}z_\mathrm{t}\approx const. 
\label{eq-parat}
\eq
which we use to maintain dynamical equilibrium, when changing the thick disc parameters.
For the stellar population of the thick disc we adopt a simple population with an age of 12\,Gyr
and metallicity [Fe/H]=-0.7.

We include a simple stellar halo described by a flattened power law distribution
\bq
\rho_\mathrm{h} = \rho_\mathrm{h,0}
	\left(\frac{R^2+z^2/b^2}{R_0^2}\right)^{\alpha_\mathrm{h}/2}
\label{eq-rhoh}
\eq
with flattening $b$ and local normalisation $\rho_\mathrm{h,0}$. For the distance of the Sun to the Galactic centre we adopt $R_0=8.0$\,kpc. Since we are looking only to one line of sight and to
distances not large compared to $R_0$, the parameters $\alpha_\mathrm{h}$ and
$b$ are strongly degenerated. A stronger flattening requires a shallower slope
in order to reproduce similar stellar densities at a distance of $z=5- 10$\,kpc. One choice to get minimum $\chi^2$ values is
$\alpha_\mathrm{h}=-3.0$ and $b=0.7$, which we fix for the further investigations.

For the calculation of star counts we use the density profiles of each component
 $\nu$ at distance $s$ along the line of sight 
pointing to the Galactic coordinate position $(l,b)$ 
normalised at the solar position $(R,z)_{\sun}=(R_0,0)$ 
\bqn
\rho_\mathrm{\nu}(s)&=&\frac{\rho(R,z)}{\rho_0}\quad\mbox{with}\quad
\rho_0=\rho(R_0,0),\\
 z=s\sin b,&& R=\sqrt{R_0^2-2R_0s\cos b\cos l+s^2\cos^2 b}.
\label{eq-rhonu}
\eqn
For the thin disc the density profile depends also on colour $g-r$.
The vertical offset $z_0\approx 20$\,pc of the solar position can be easily included but is
negligible for high Galactic latitude fields.
The density distributions of thin and thick disc can be
extended in the radial direction by an exponential profile, if necessary.
The normalised density profiles of each component are transformed to number density profiles
 by multiplying with the local number densities $n_{\nu,0}(g-r,M_\mathrm{g})$ 
 determined from the  HRD in the solar neighbourhood.

\section{Local Normalisation}\label{sec-local}

The Galaxy model described in the last section must be complemented by the
local stellar number density $n_{\nu,0}(g-r,M_\mathrm{g})$ of each component $\nu$. 
In general the full HRD of a volume complete sample should be used. Since most
of the modelled CMD is dominated to more than 95 percent by MS stars, 
we restrict the star count predictions to MS stars including a correction for
turnoff stars. Only at the bright red corner of the CMD the K
giants of thick disc and halo contribute significantly.

The stellar content in the solar neighbourhood is quantified by
the number of MS stars $N_{25}$ with 
\bq
N_{25}=\sum_{\nu}N_{25,\nu}(g-r,M_\mathrm{g}) \quad\mbox{with} \quad
\nu=\mbox{s,t,h}
\eq
in a sphere of 25\,pc radius with volume $V_{25}=65,450$\,pc$^3$.
Here the contributions of thin disc, thick disc, and halo are denoted by the
indices s, t, and h, respectively.
In \citet{Ju08} absolute magnitudes and colours of the mean MS in the 
$u,g,r,i,z$ filter system were determined based on different
transformation formulae available in the literature. We use that mean MS for the
thin disc, because the influence of thick disc and halo stars in the solar
neighbourhood is negligible.
\begin{figure}
\centerline{\resizebox{0.98\hsize}{!}{\includegraphics[angle=270]{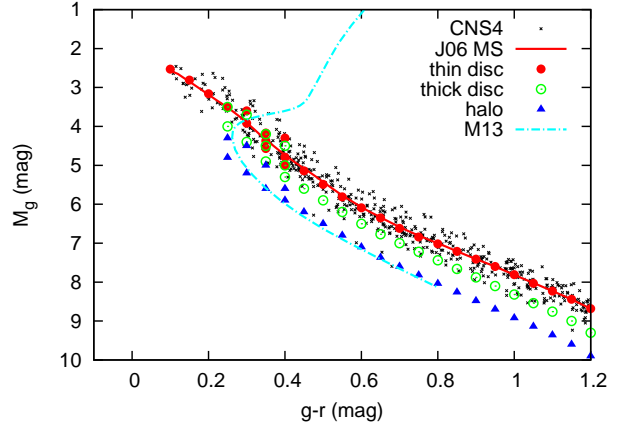}}}
\caption[]{
The local MS of thin disc (red full circles and line), thick disc (green open circles), and stellar
halo (blue triangles). The MS stars of the CNS4  and the fiducial
sequence of M13 with metallicity [Fe/H]=-1.53 are over-plotted.
}
\label{fighrd}
\end{figure}

In figure \ref{fighrd} we show $M_\mathrm{g}(g-r)$ for the thin disc 
based on the transformation J06 \citep{Jo06}. The black crosses denote the MS stars of the 
CNS4 which are the basis for the determination of the mean MS.
Each red full circle along the MS denotes a data
point, where $N_\mathrm{25,s}(g-r,M_\mathrm{g})$ is a free fitting parameter. 
The locus of the MS
corresponds to a slightly sub-solar mean metallicity and is consistent with the
fiducial sequences of the open clusters M\,67 ([Fe/H]=0) and NGC\,2420 ([Fe/H]=-0.37) of \citet{An08}.
Since the
mean metallicity of the thin disc population decreases with increasing distance
from the mid-plane, a correction to the mean luminosity in each colour bin is
necessary. We use a simple analytic approximation 
\bq
\Delta M_\mathrm{g}([Fe/H](z)) = 0.4\left(\frac{z}{z+600\mathrm{pc}}\right)^2\mathrm{mag},
\label{eq-DeltaMg}
\eq
which is consistent with the determination of \citet{Iv08} for the $M_{r}$
 luminosity of thin disc MS stars.

For the thick disc we use an interpolated MS with 
$N_\mathrm{25,t}(g-r,M_\mathrm{g})$  corresponding to an old population
with metallicity 
[Fe/H]\,$\approx-0.7$ (green open circles in figure \ref{fighrd}). 

The stellar halo is represented by $N_\mathrm{25,h}(g-r,M_\mathrm{g})$ corresponding
to an old population with metallicity
[Fe/H]$=-1.5$ (blue triangles in figure \ref{fighrd}), where we used the
fiducial sequence of M15 \citep{An08} with an extrapolation to the faint end. 

In the turnoff regime of F stars we split the MS luminosity into two or three
points with different luminosities in order to include the brighter turnoff
stars. The total number of thin disc, thick disc and halo should add up to the
observed total number
of MS stars $N_{25}(g-r)$, which were also determined in \citet{Ju08}
(histogram in figure \ref{fign-gr}).  
We treat the
local number densities of the components as free fitting parameters
and then compare the results
of the best-fitting model to the observed
stellar content in the solar neighbourhood.

The predicted Hess diagrams, i.e. the star counts in the CMD $N(g-r,g)$ per 
$\dd(g-r)\dd g$ bin, in a cone with cross section $\dd l\cos b \dd b$ 
pointing to the Galactic coordinate position $(l,b)$ are calculated 
by adding up the contributions of each component along the line of sight. 
The contribution of each component $\nu$ by a volume element at distance $s$, using 
$\dd s/s=0.2\log(10)\,\dd g$, is given by
\bqn
\Delta N_{\nu}\dd g &=&N_{25,\nu}(g-r,M_\mathrm{g})\rho_\mathrm{\nu}(s) 
	s^2\dd s\dd l\cos b \dd b \label{eq-dN}\\
&=&\frac{N_{25,\nu}(g-r,M_\mathrm{g})}{V_{25}}
	\frac{\rho_\mathrm{\nu}(s)s^3\dd g}{2.17}
	\frac{4\pi}{41,253\deg^2}\nonumber\\
g &=& M_\mathrm{g}+\Delta M_\mathrm{g}+5\log_{10} s -5 .
\nonumber
\label{eq-DeltaN}
\nonumber
\eqn
Adding up the contributions of all components along the line of sight yields the
predicted number $N(g-r,g)$ in each bin of the Hess diagram. In the next step the values
are smoothed in the same way as the SDSS data.
The star numbers in the Hess diagrams are normalised to 1\,deg$^2$ at the sky
(second line of equation \ref{eq-dN}).
In $(g-r,g)$ we need to distinguish between the resolution $\dd(g-r)\dd g$, the
smoothing $\Delta g$ and the
normalisation. All Hess diagrams are normalised to 
0.1\,mag in $(g-r)$ and 1.0\,mag in $g$.
\begin{figure}
\centerline{\resizebox{0.67\hsize}{!}{\includegraphics[angle=0]
{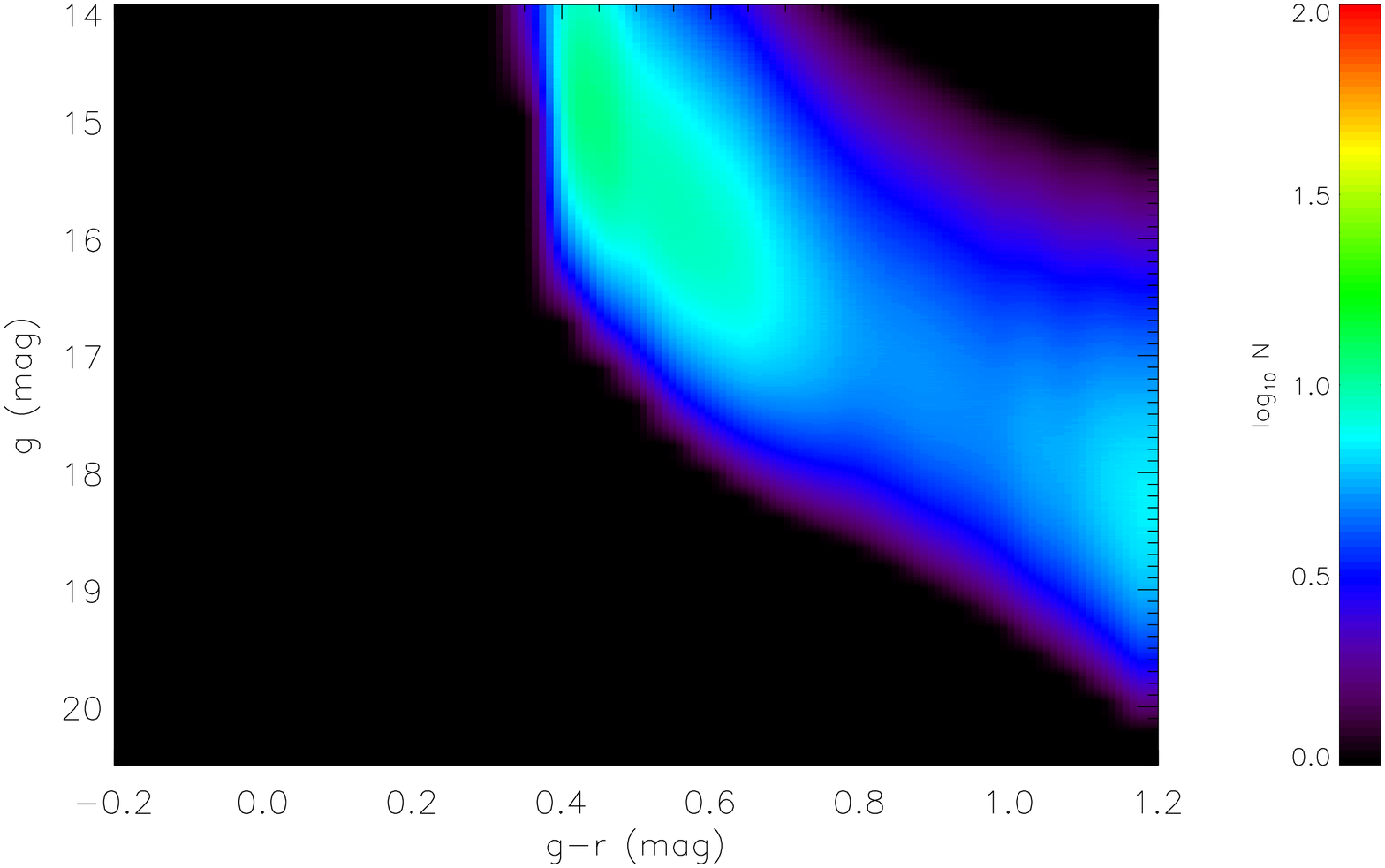}}}
\centerline{\resizebox{0.67\hsize}{!}{\includegraphics[angle=0]
{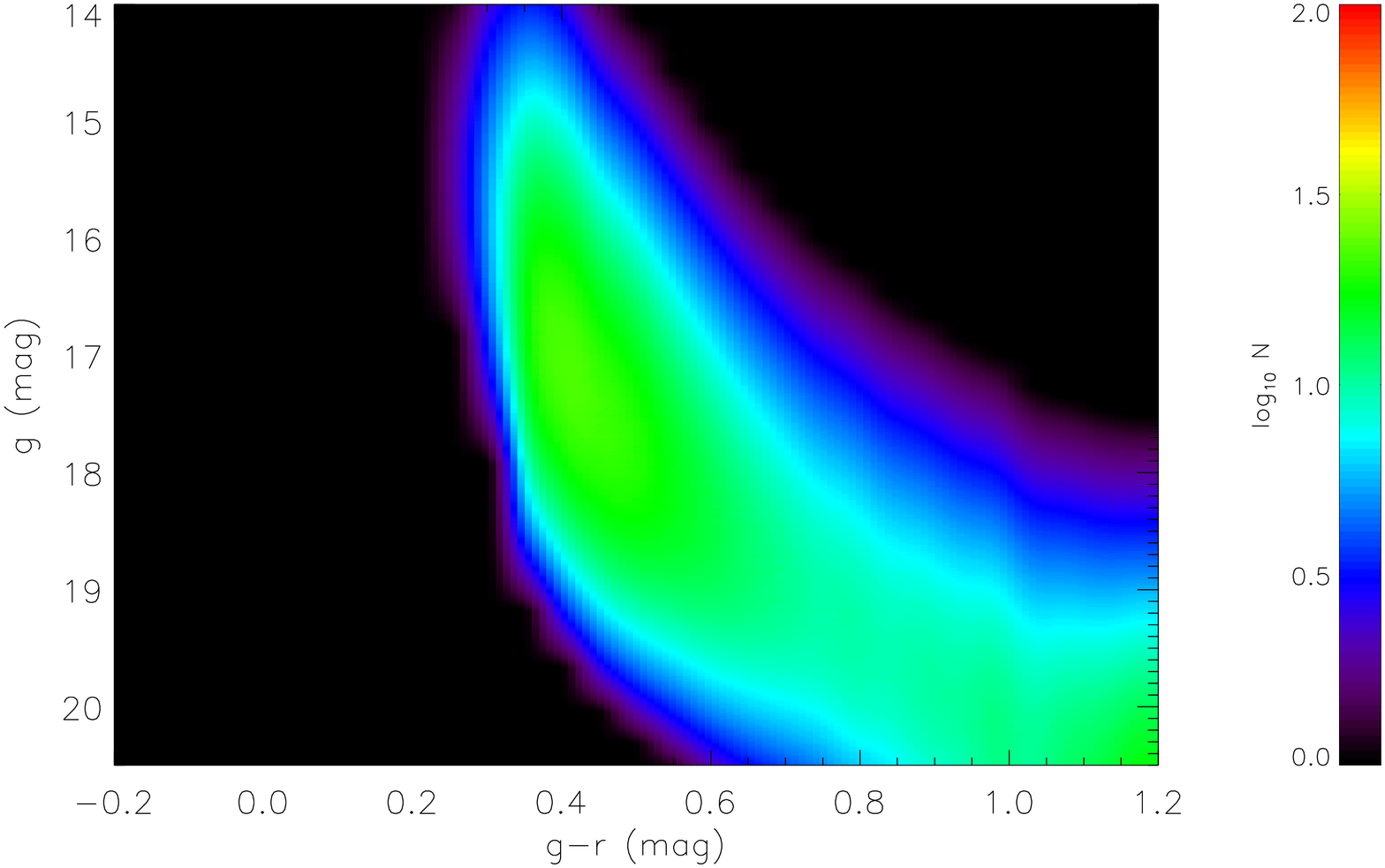}}}
\centerline{\resizebox{0.67\hsize}{!}{\includegraphics[angle=0]
{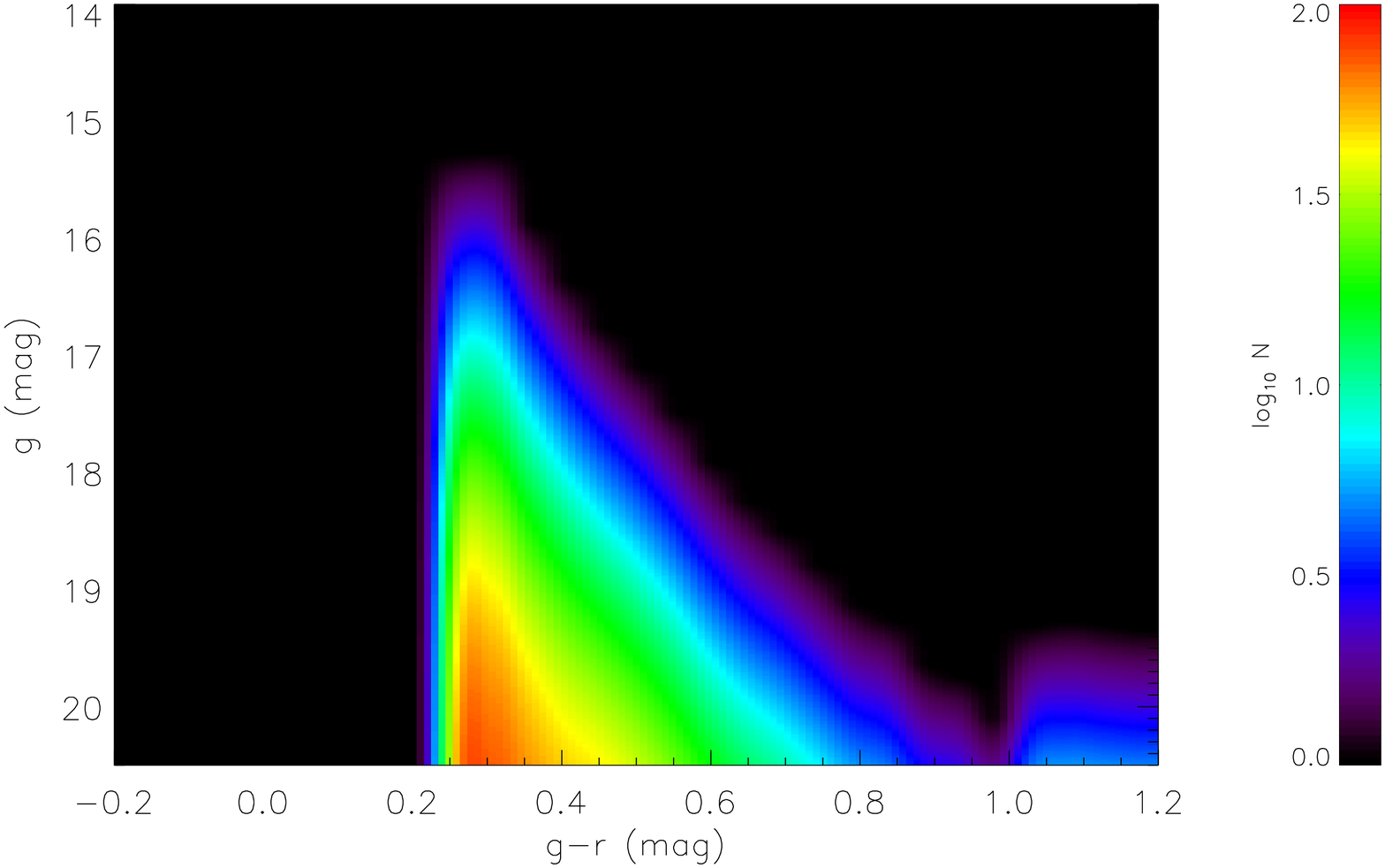}}}
\centerline{\resizebox{0.67\hsize}{!}{\includegraphics[angle=0]
{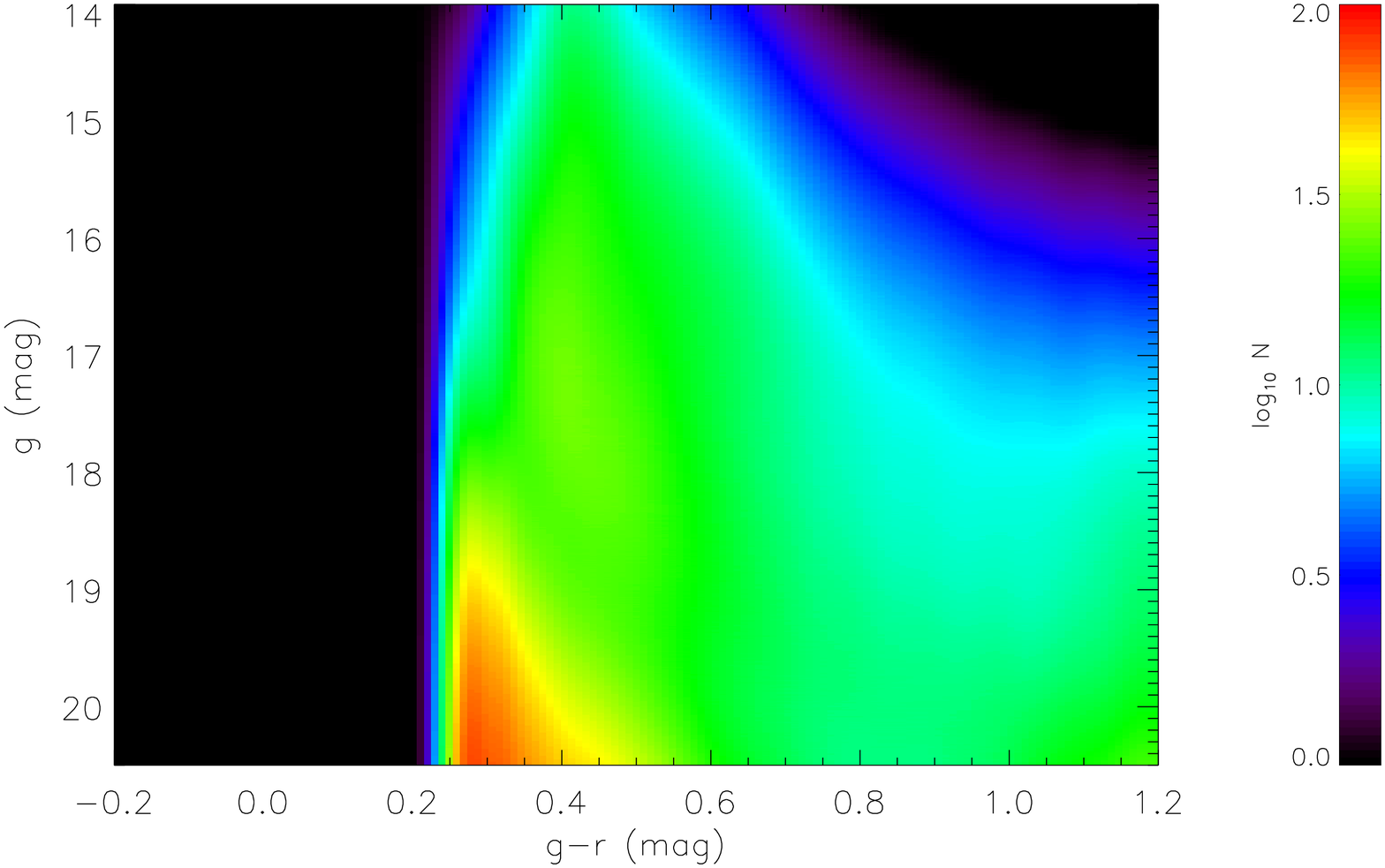}}}
\centerline{\resizebox{0.67\hsize}{!}{\includegraphics[angle=0]
{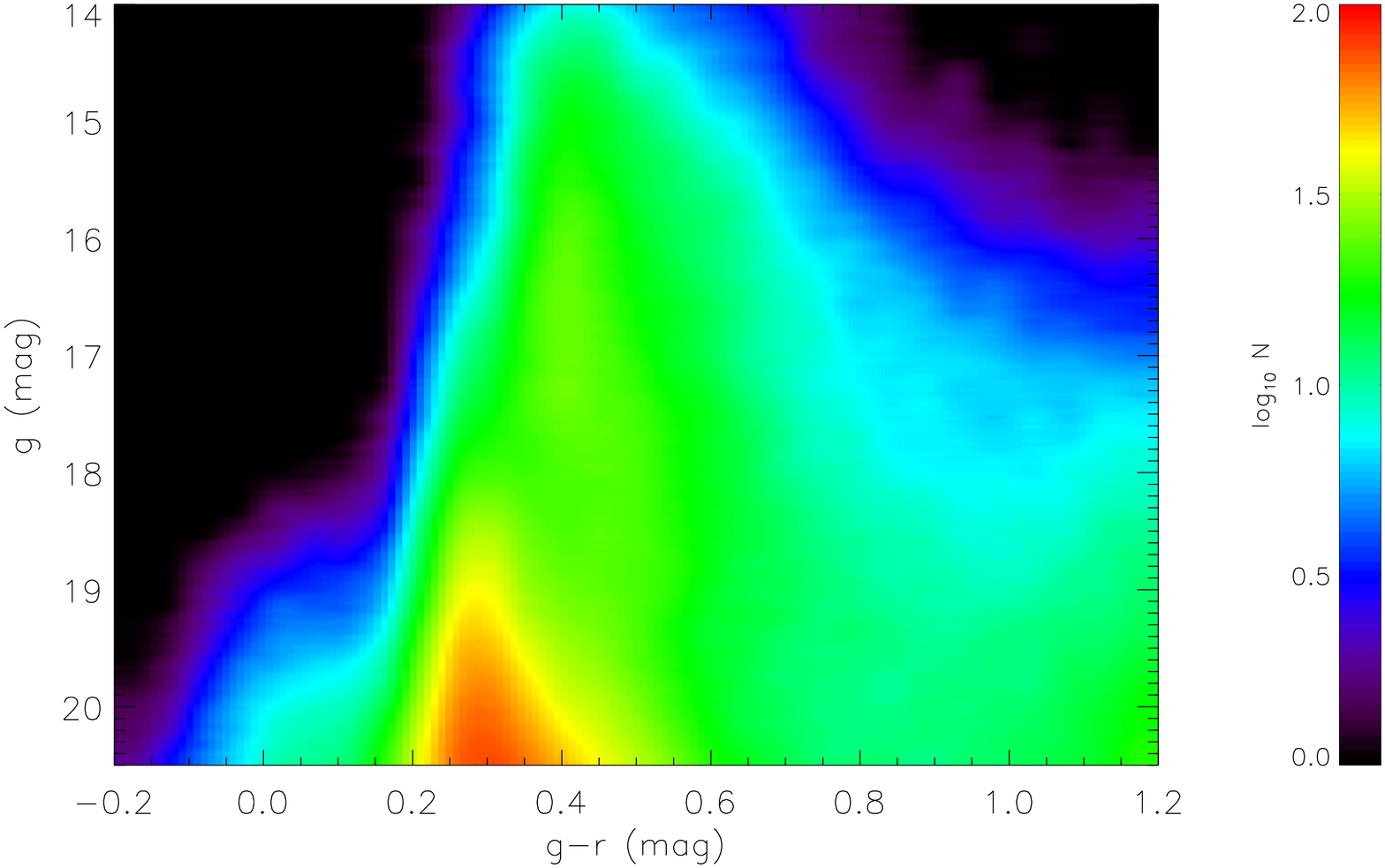}}}
\caption[]{
Top to bottom: Contributions of thin disc, thick disc stellar
halo, and full Hess diagram of model A-R06-05 (see table \ref{tab-fit})
 followed by the NGP data. Colour coding
ranges from 1 (purple) to 100/deg$^2$ (red) in log-scale.
}
\label{figcomp}
\end{figure}

\section{Fitting procedure}\label{sec-fit}

The best-fit procedure in the colour bins are independent of each other. In
each colour bin the
contribution of thin disc, thick disc, and stellar halo to the local star
counts in $V_{25}$ are
quantified by the free parameters $N_\mathrm{25,s}$, $N_\mathrm{25,t}$
and $N_\mathrm{25,h}$, respectively. The $(g-r)$ arguments are
dropped here for simplicity. 
We use the nonlinear Levenberg-Marquard algorithm \citep{Pr92} to minimise
in each colour bin $i$
\bq
\chi^2_\mathrm{i}=\sum_{g_\mathrm{j}}
\frac{(\log Y_\mathrm{ij} - \log N_\mathrm{ij})^2}{\sigma_\mathrm{ij}^2}
\quad\mbox{with}\quad \sigma_\mathrm{ij}^2=\frac{1}{Y_\mathrm{ij}}
\label{eq-chi2tot}
\eq
in log-scale with Poisson noise for the statistical weights $\sigma_\mathrm{ij}$.
 In linear
scale the regions with maximum density would dominate the $\chi^2$ value resulting
in an unsatisfactory overall fit.
The $\chi^2_\mathrm{i}$ are derived from normalised values and have to be
corrected by the area $A_\mathrm{NGP}$ in units of deg$^2$ and the
bin size $\dd (g-r)\Delta g$ in units of 
$0.1\,\mathrm{mag}\times 1\,\mathrm{mag}$. 
Due  to the smoothing in $g$ over $n$ data points $\chi^2_\mathrm{i}/n$ is the
average over all subsets of independent bins with $\chi^2_\mathrm{i,k}$ shifted
by $n\times\dd g$. 
In each colour bin the degrees of freedom $dof_\mathrm{i}$ for the independent bin sets depend on
the fit regime in $g$, the bin size $\Delta g$ and the number of fitting
parameters. We derive the mean reduced $\chi^2$ by adding up the
normalised $\chi^2_\mathrm{i}$ values
\bq
\chi^2=\frac{A_\mathrm{NGP}}{1\,\mathrm{deg}^2}
\frac{\dd (g-r)}{0.1\,\mathrm{mag}}\frac{\Delta g}{1\,\mathrm{mag}}
\sum_i\frac{\chi^2_\mathrm{i}}{n\times dof_\mathrm{i}}.
\label{eq-chi2red}
\eq
In order to get reliable fits we restrict the colour range of the fitting regime.
For the thin disc we fix $N_\mathrm{25,s}(g-r,M_\mathrm{g})$ for $g-r< 0.35$, since
the main contribution falls outside the bright limit $g=14$\,mag.
For the halo we extrapolate $N_\mathrm{25,h}(g-r,M_\mathrm{g})$ for $g-r> 1.0$, since
the main contribution falls outside the faint limit $g=20.5$\,mag.
Additionally the part of the CMD, which we use to minimise $\chi^2$ is
restricted dependent on the aspect of investigation.

\section{Results}\label{sec-results}

We discuss first the construction of the Hess diagrams and compare the best fit
results of models A -- D. Then we investigate the dependence of the fitting result
on the fitting regime and the MS properties for model A.

\subsection{Hess diagram fitting}

The bottom panel of figure \ref{figcomp} shows the Hess diagram of the NGP data
with a total of 276,180 stellar objects. Star number densities normalised to 
0.1\,mag$\times$1.0\,mag in $(g-r,g)$ and 1\,deg$^2$ are colour coded in
log-scale ranging from 1\dots100 (purple \dots red). The plots are smoothed in
colour and magnitude in boxes in steps of 0.01\,mag with box size $\Delta
(g-r)\times\Delta g=$0.05\,mag$\times$0.5\,mag (in the fitting procedure the
colour bins are not smoothed). Most of the Hess diagram is strongly dominated by
MS stars. Exceptions are turn-off stars in the colour range
$g-r\approx$\,0.3--0.4 and red giants of thick disc and/or halo in the upper right
corner. The nature of the faint very blue objects with $g-r<$0.15 is unclear
(miss-identified extragalactic sources, White Dwarfs, halo BHB stars, \dots).

\begin{figure*}
\centerline{\resizebox{0.3\hsize}{!}{\includegraphics[angle=0]
{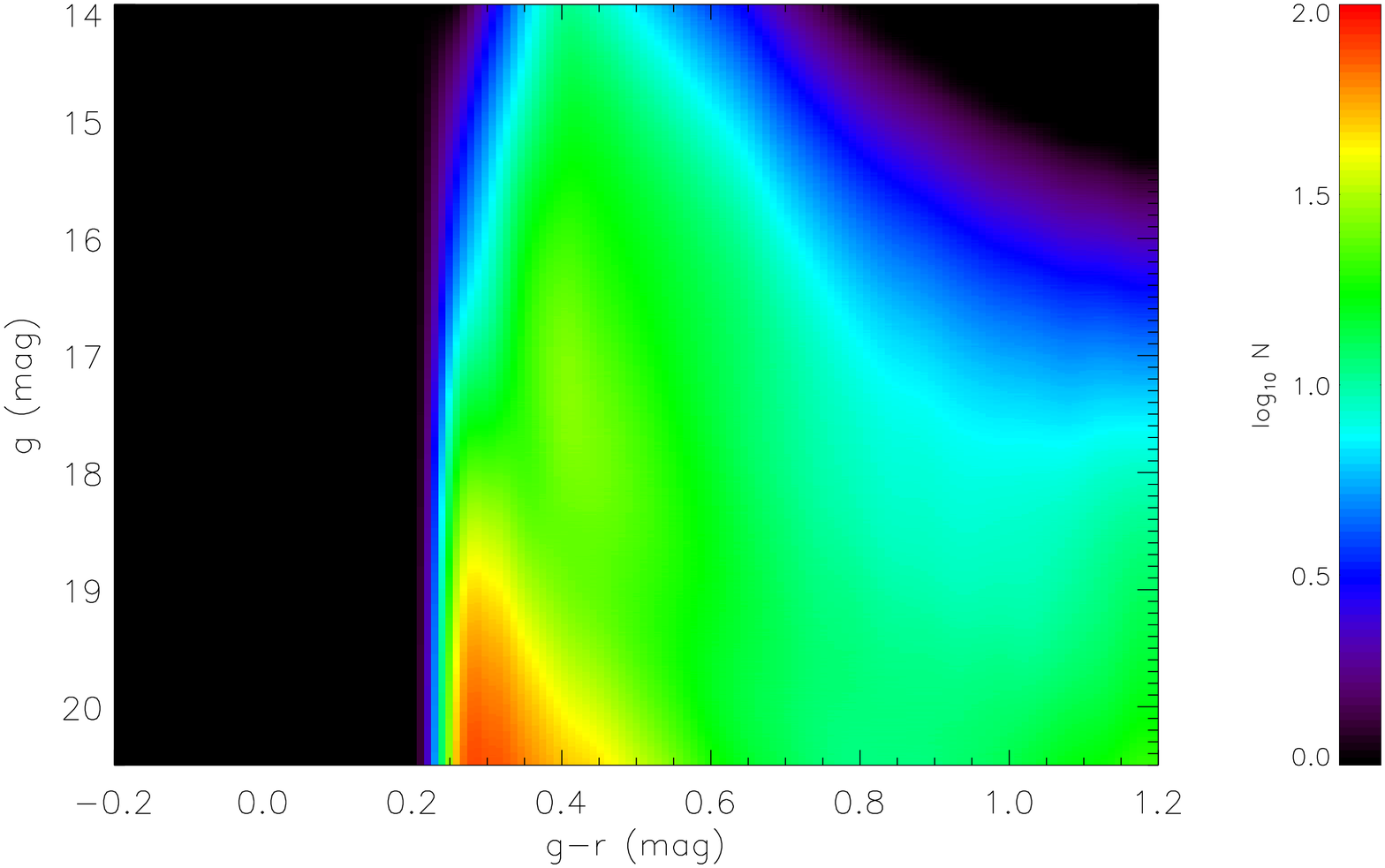}}
\resizebox{0.3\hsize}{!}{\includegraphics[angle=0]
{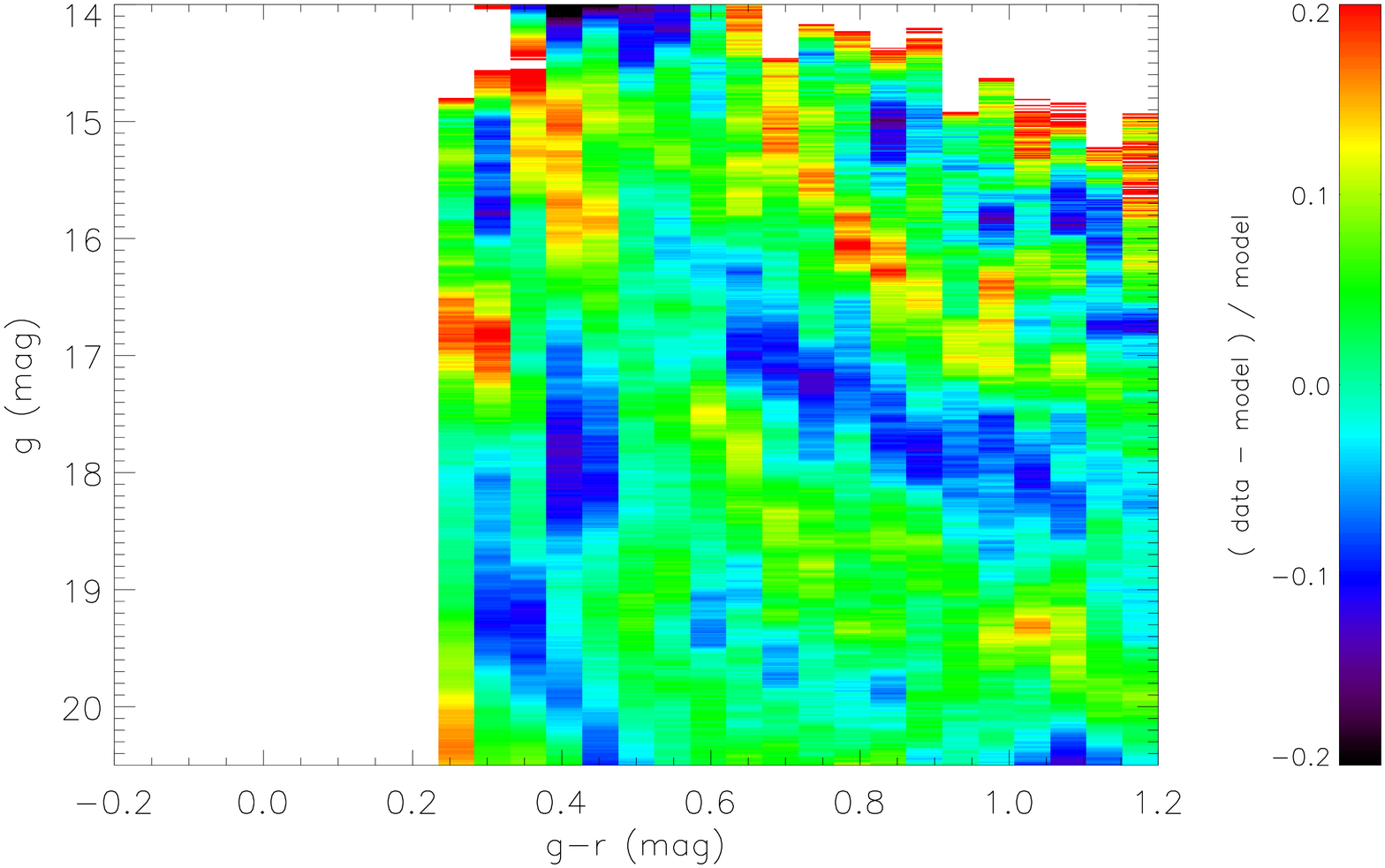}}
\resizebox{0.3\hsize}{!}{\includegraphics[angle=0]
{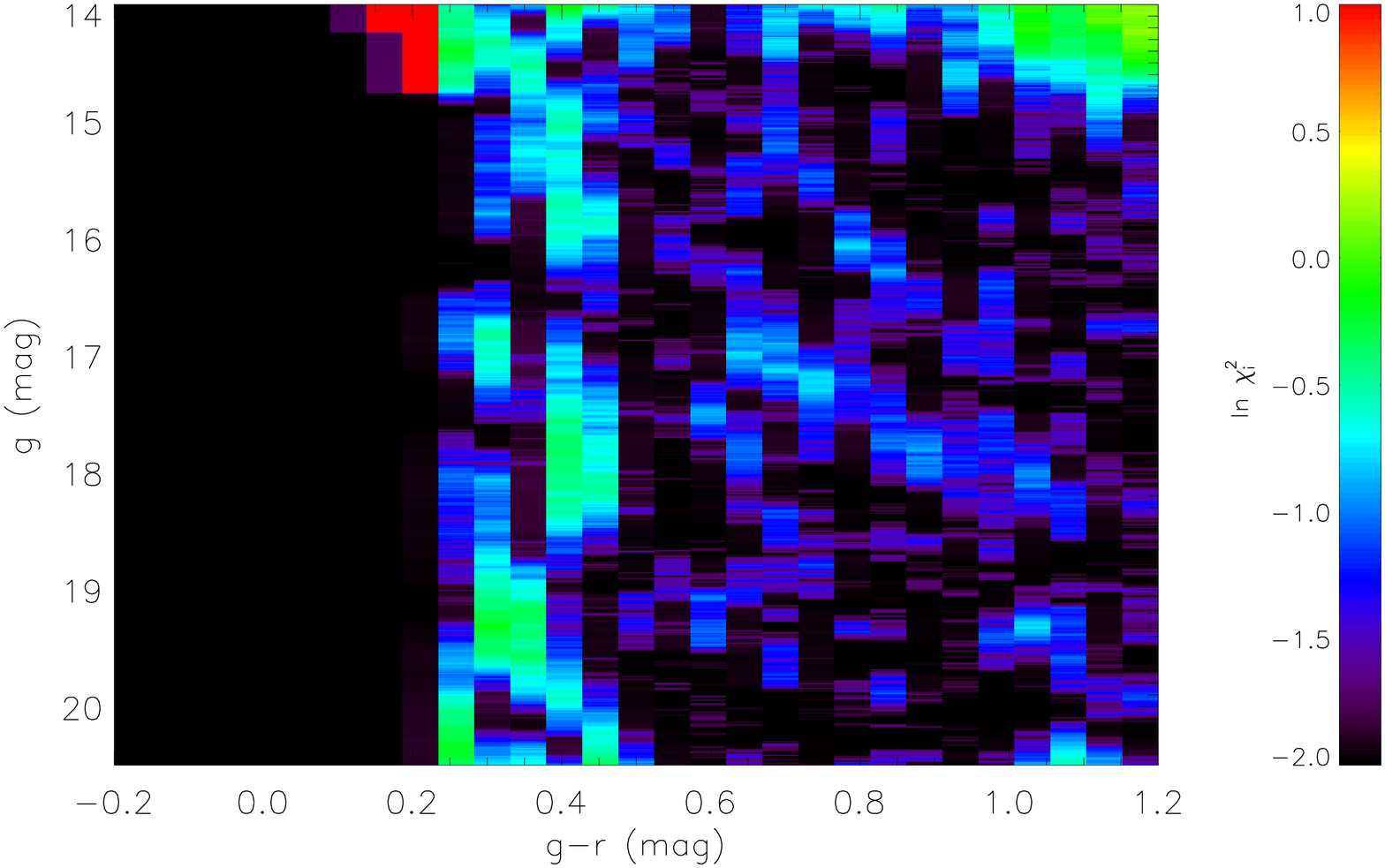}}}
\centerline{\resizebox{0.3\hsize}{!}{\includegraphics[angle=0]
{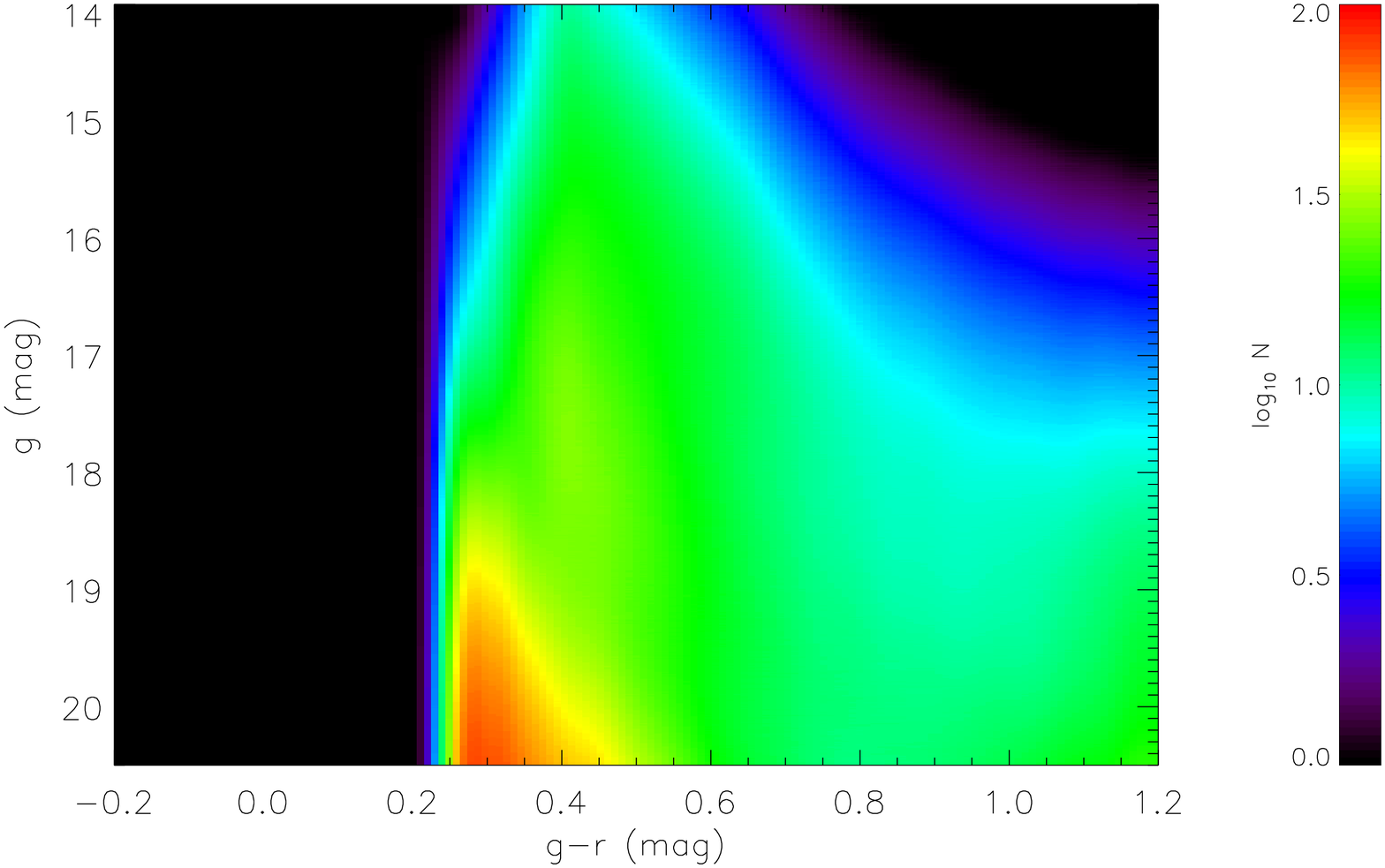}}
\resizebox{0.3\hsize}{!}{\includegraphics[angle=0]
{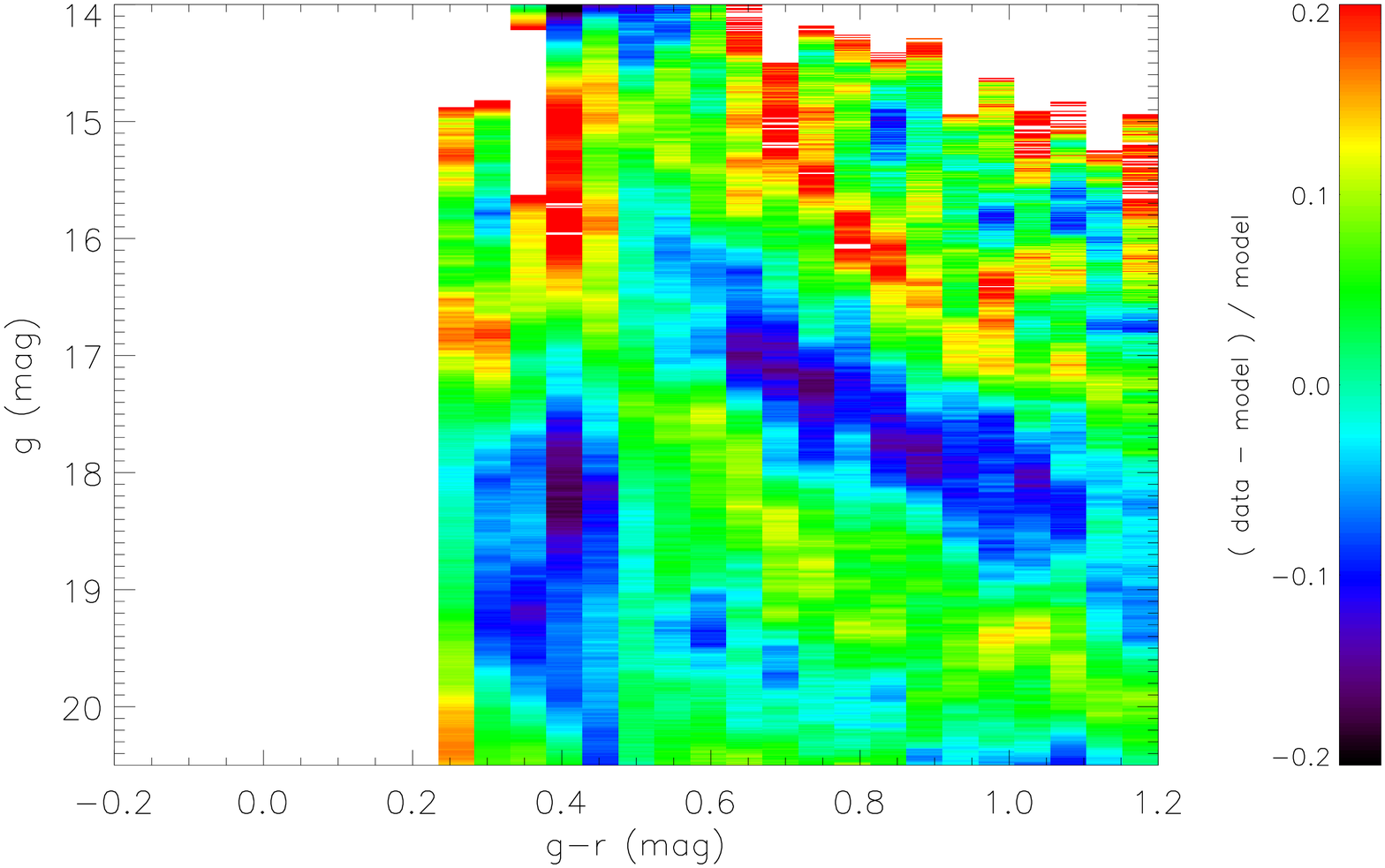}}
\resizebox{0.3\hsize}{!}{\includegraphics[angle=0]
{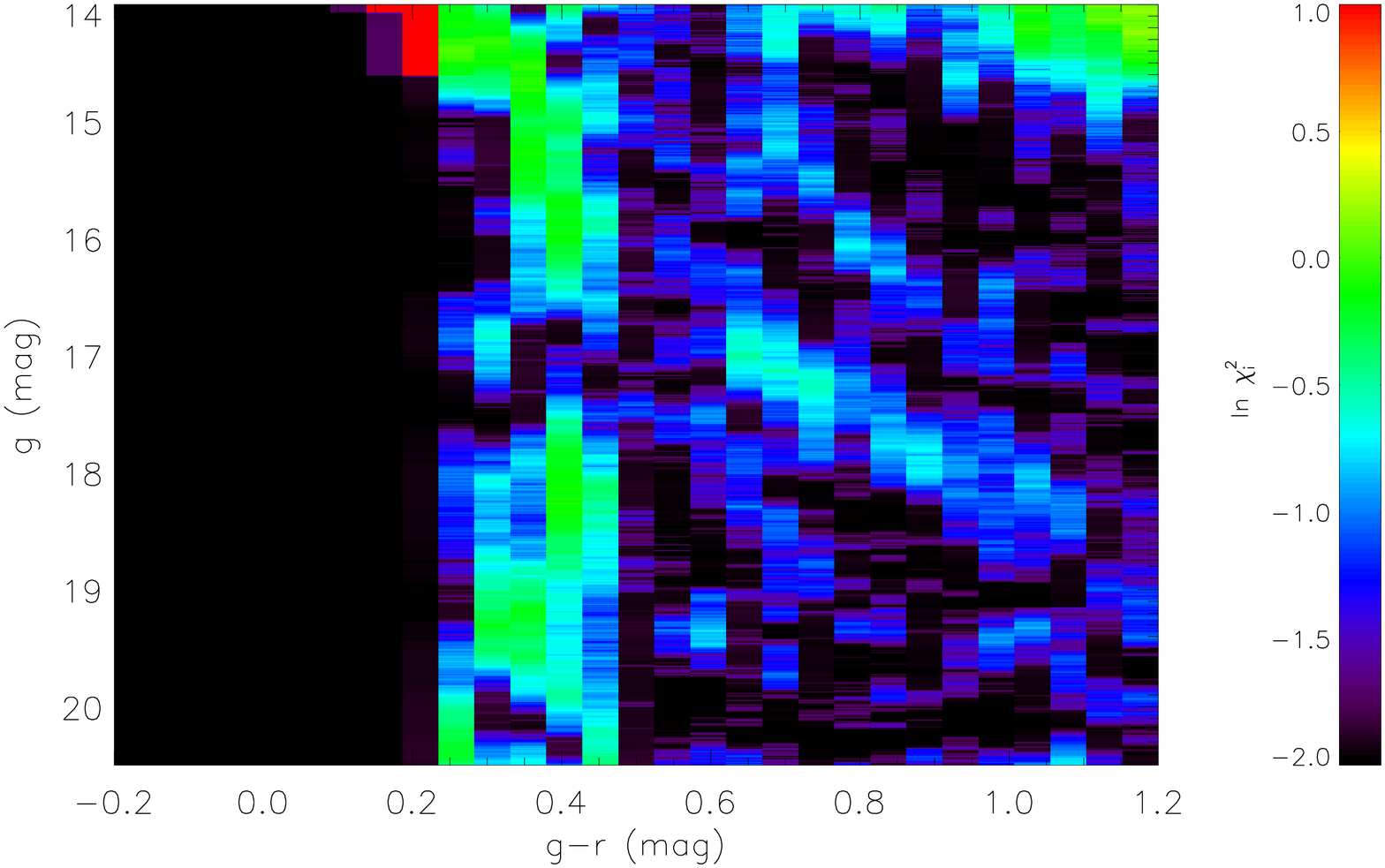}}}
\centerline{\resizebox{0.3\hsize}{!}{\includegraphics[angle=0]
{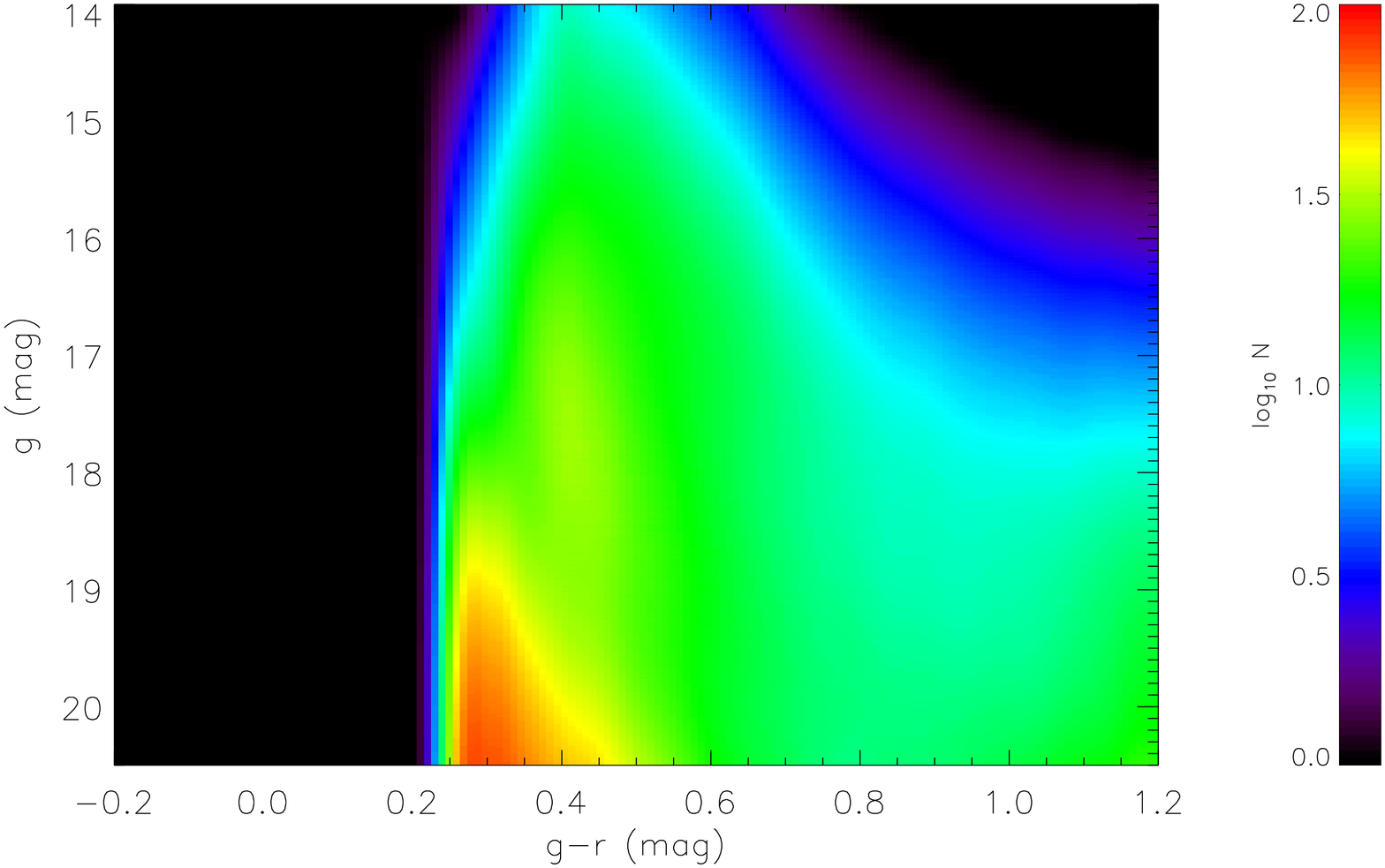}}
\resizebox{0.3\hsize}{!}{\includegraphics[angle=0]
{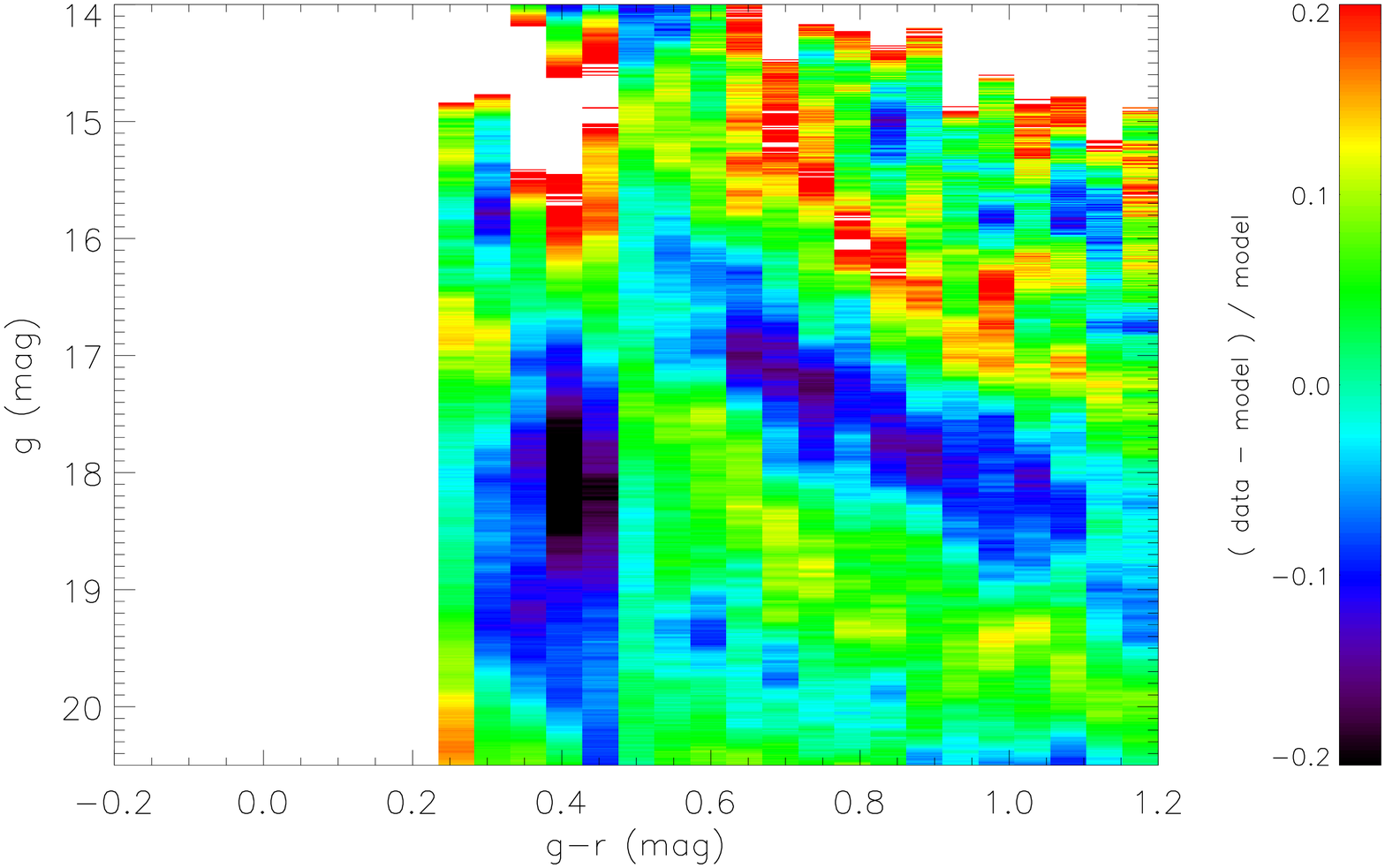}}
\resizebox{0.3\hsize}{!}{\includegraphics[angle=0]
{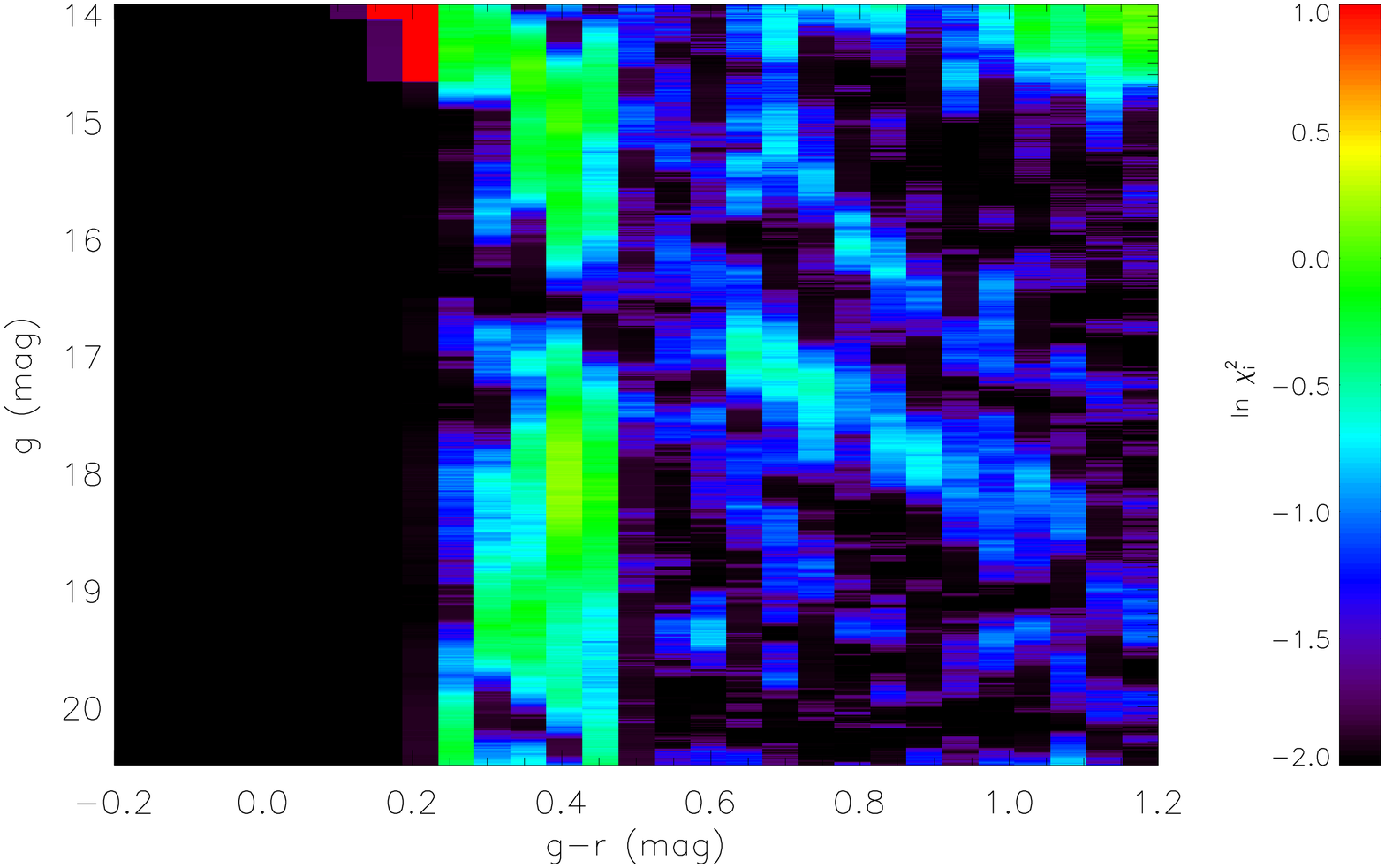}}}
\centerline{\resizebox{0.3\hsize}{!}{\includegraphics[angle=0]
{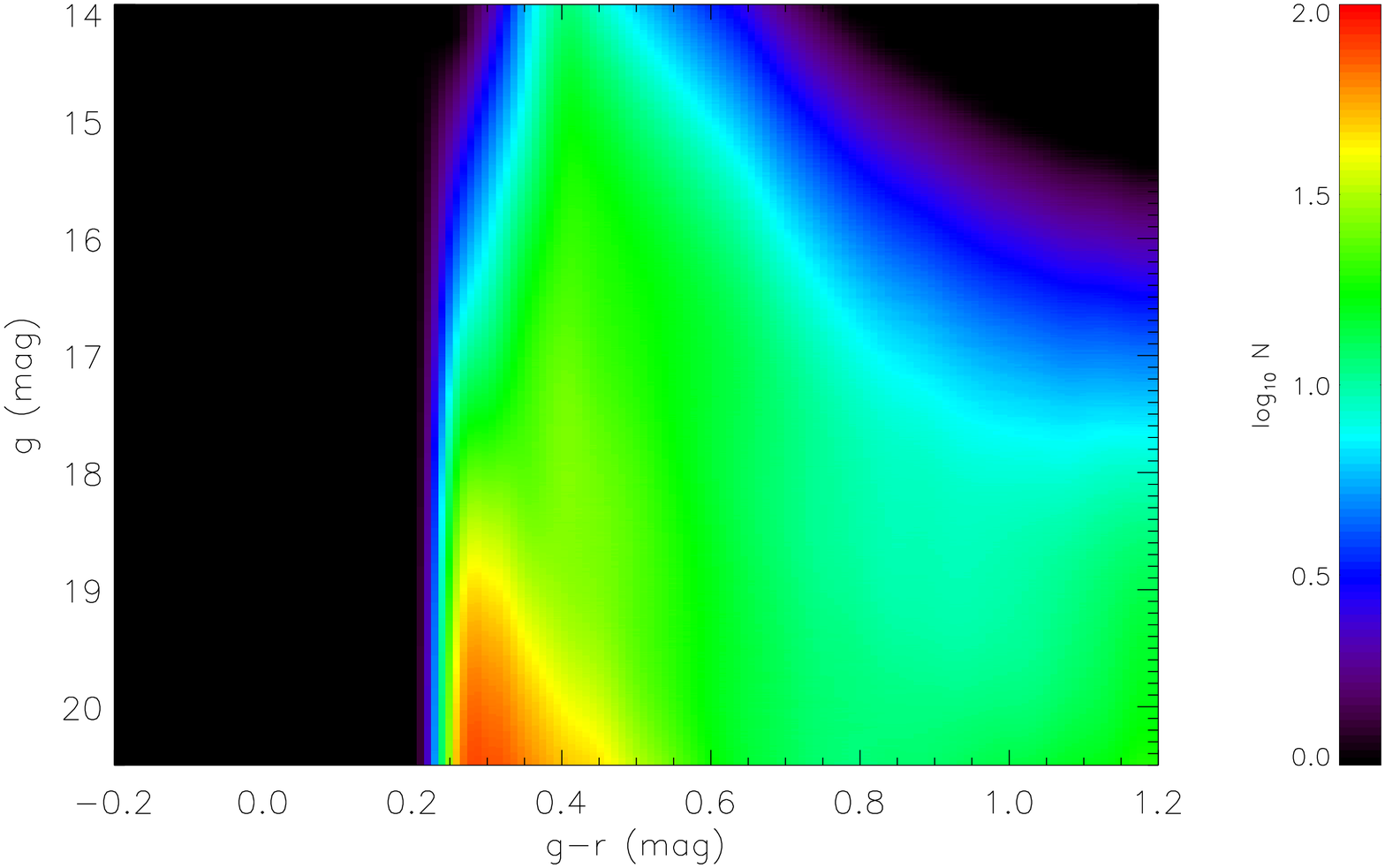}}
\resizebox{0.3\hsize}{!}{\includegraphics[angle=0]
{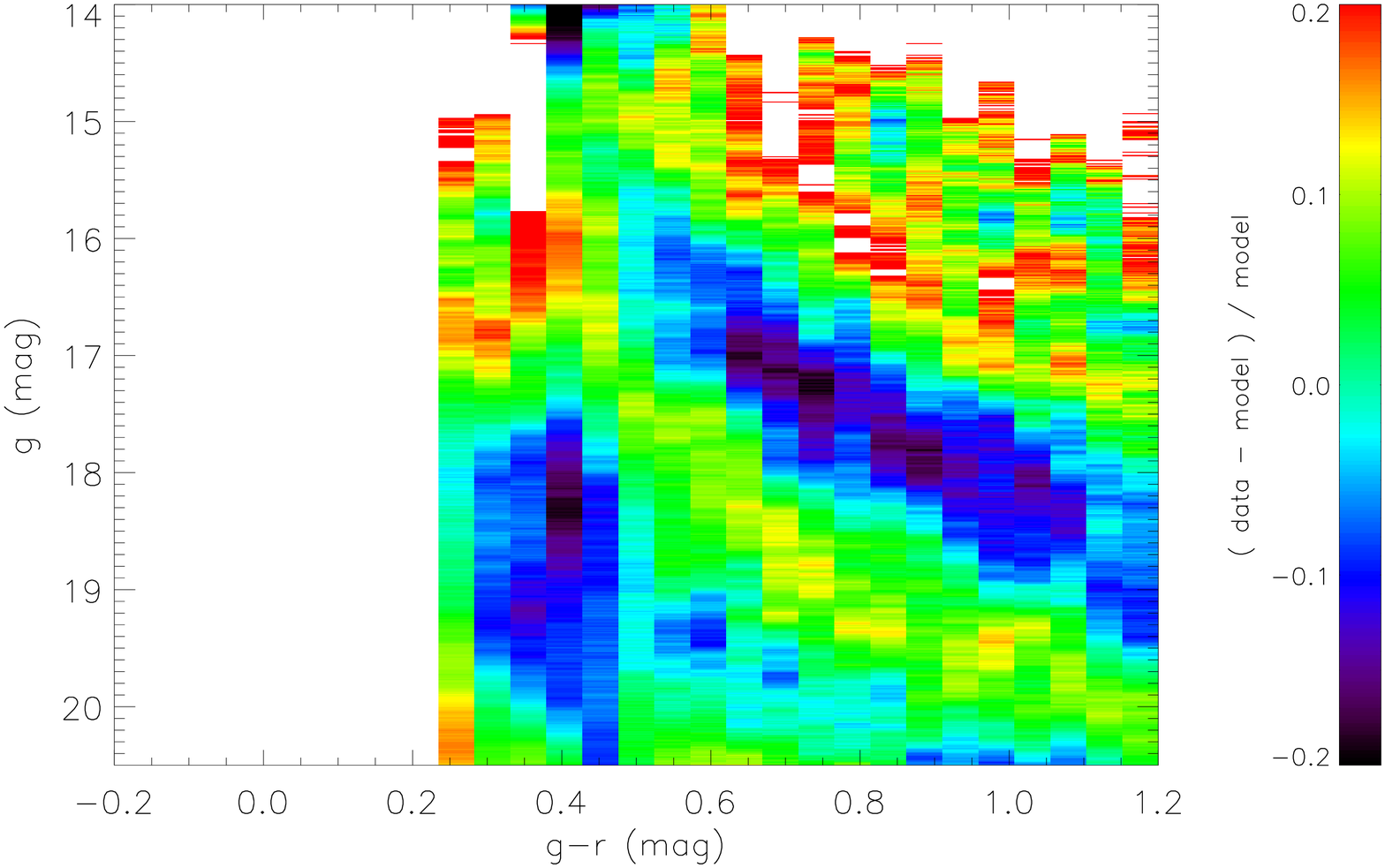}}
\resizebox{0.3\hsize}{!}{\includegraphics[angle=0]
{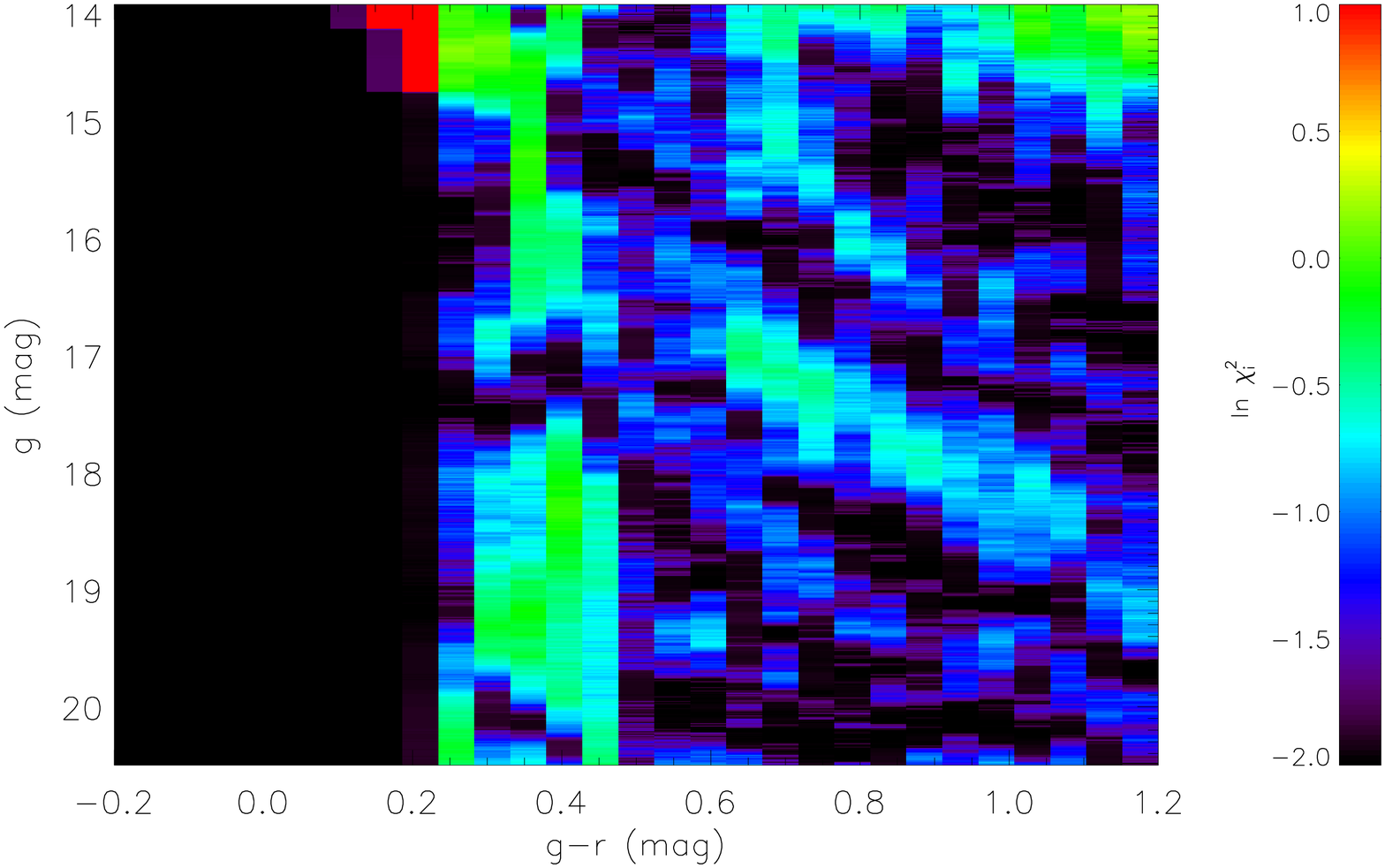}}
}
\caption[]{
Comparison of models A,B,C,D-J06-05-r (top to bottom). 
From left to right: Hess diagram (same colour coding as in figure
\ref{figcomp}), relative difference to SDSS data ranging from -0.2 (purple) to
+0.2 (red) in linear scale, $\chi^2_\mathrm{ij}$ distribution in log-scale.
}
\label{figA-D}
\end{figure*}

The three top panels of figure \ref{figcomp} show the contributions of the different
components to the full Hess diagram for model A. There is a significant overlap
of thin/thick disc, as well as thick disc/halo. Therefore the stellar halo must
be included to determine the thick disc properties and the thick disc is needed
to fix the thin disc parameters. The second last plot shows the sum of all
components for model A.

For the comparison of disc models A -- D we restrict the colour range to
$0.5\le g-r\le 1.2$\,mag in order to avoid the F turn-off regime, which may be
improperly modelled and thus dominate the total $\chi^2$. The 
magnitude range is $14.5\le g\le 20.5$\,mag for safely excluding a contribution of
saturated stars brighter than 14.25\,mag in the brightest bins and a significant
contamination by misidentified extragalactic sources at the faint end. In all models the
thick disc parameters are varied according to equation \ref{eq-parat} to
find the minimum $\chi^2$ value. 
The parameters of the best fits are collected in the first four rows of table
\ref{tab-fit}. In the model names there are tokens for the local
normalisation (-J06 for \citet{Jo06}, -R06 for \citet{Ro06}, -C08 for \citet{Ch08})
and the smoothing $\Delta g$ (-01, -02, -05, respectively) attached. 
For the first block '-r' is attached for the reduced fit regime.
\begin{table}
\caption{Parameters of the best fit solutions.}
\begin{tabular}{c c c c c c c c c c c c}
\hline
model & $\Delta g [mag]$ & $\chi^2$ & $N_\mathrm{25}$ & $z_\mathrm{t} [pc]$ &
 $\alpha_\mathrm{t}$ & $\sigma_\mathrm{t}[km/s]$\\
A-J06-05-r& 0.5       & 2.59 & 739   &  800  & 1.16    & 45.3 \\ 
B-J06-05-r& 0.5       & 3.75 & 748   &  880  & 1.07    & 47.4 \\ 
C-J06-05-r& 0.5       & 3.84 & 811   &  885  & 1.23    & 51.2 \\ 
D-J06-05-r& 0.5       & 5.64 & 709   &  930  & 0.99    & 48.0 \\ 
A-R06-05-r& 0.5       & 2.61 & 715   &  800  & 1.16    & 45.3 \\ 
A-C08-05-r& 0.5       & 2.60 & 733   &  800  & 1.16    & 45.3 \\ 
\hline
A-J06-05& 0.5  & 4.50 & 762   &  800  & 1.16	& 45.3 \\ 
A-R06-05& 0.5  & 4.31 & 710   &  800  & 1.16	& 45.3 \\ 
A-C08-05& 0.5  & 5.09 & 752   &  800  & 1.16	& 45.3 \\ 
A-C08-02& 0.2  & 2.53 & 750   &  800  & 1.16	& 45.3 \\ 
A-C08-01& 0.1  & 1.84 & 753   &  800  & 1.16	& 45.3 \\ 
\hline
\end{tabular}

{\it Note.} The models are named by the disc model A -- D followed by the local
normalisation and the smoothing $\Delta g$. For the first block 'r' is added for the
reduced fit regime. The observed number of stars range from
 $N_\mathrm{25}=726$ (with R06) to $N_\mathrm{25}=770$ (with C08). 
\label{tab-fit}
\end{table}

\subsection{The star formation history}

The thin disc models A -- D have very different SFRs leading to different star count
predictions. In each model we optimised the thick disc scale height and power law index according to equation \ref{eq-parat} starting from a self-consistent
isothermal model. In figure
\ref{figA-D} the Hess diagrams of the models are presented in the left column
(A -- D: top -- bottom). The middle
column shows the relative differences (data-model)/model, where the
coloured region covers the range -20 percent (purple) to +20 percent (red) in
linear scale. Lower values are represented by black colour and higher values
by white colour. The right column shows the contribution of each
bin to $\chi^2_\mathrm{i}$ (equation \ref{eq-chi2tot}) in log-scale.

All models fit most of the Hess diagram better than $\pm 20$\,percent. The local
normalisation of the blue
part with $g-r<0.5$, which is not included in the best fitting, is adapted by hand
to model A and not varied for the other models. In the bright red triangular
region there is a significant fraction of stars missing in the models as expected,
since the SDSS data contain red giants of thick disc and halo, which cover
that colour-magnitude regime.
 
\begin{figure}
\centerline{\resizebox{0.98\hsize}{!}
{\includegraphics[angle=270]{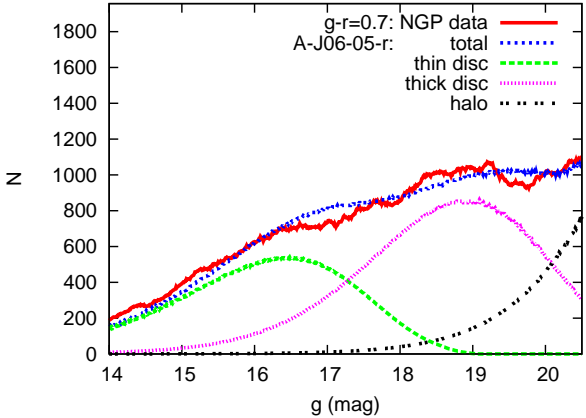}}
}
\caption[]{
Vertical cut of the Hess diagram at $g-r=0.7$\,mag. The SDSS data are compared to
the predictions of model A-J06-05-r. $N$ is the number of stars per bin 
$\dd (g-r)\Delta g= 0.05$\,mag$\times$0.5\,mag for the full NGP field. The
contributions of thin disc, thick disc, and stellar
halo are shown separately.
}
\label{fign-g}
\end{figure}

In figure \ref{fign-g} the contributions of thin disc, thick disc, and stellar
halo for model A are shown in the colour bin $g-r=0.7$ as a typical example. The maximum contribution of
the thin disc is at $g=16.5$\,mag corresponding to $z=900$\,pc and of the thick disc
at $g=19$\,mag corresponding to $z=2500$\,pc. These large distances demonstrate that it is
very important to construct models, which hold also for the outer profiles 
of the components.

\begin{figure}
\centerline{\resizebox{0.98\hsize}{!}
{\includegraphics[angle=270]{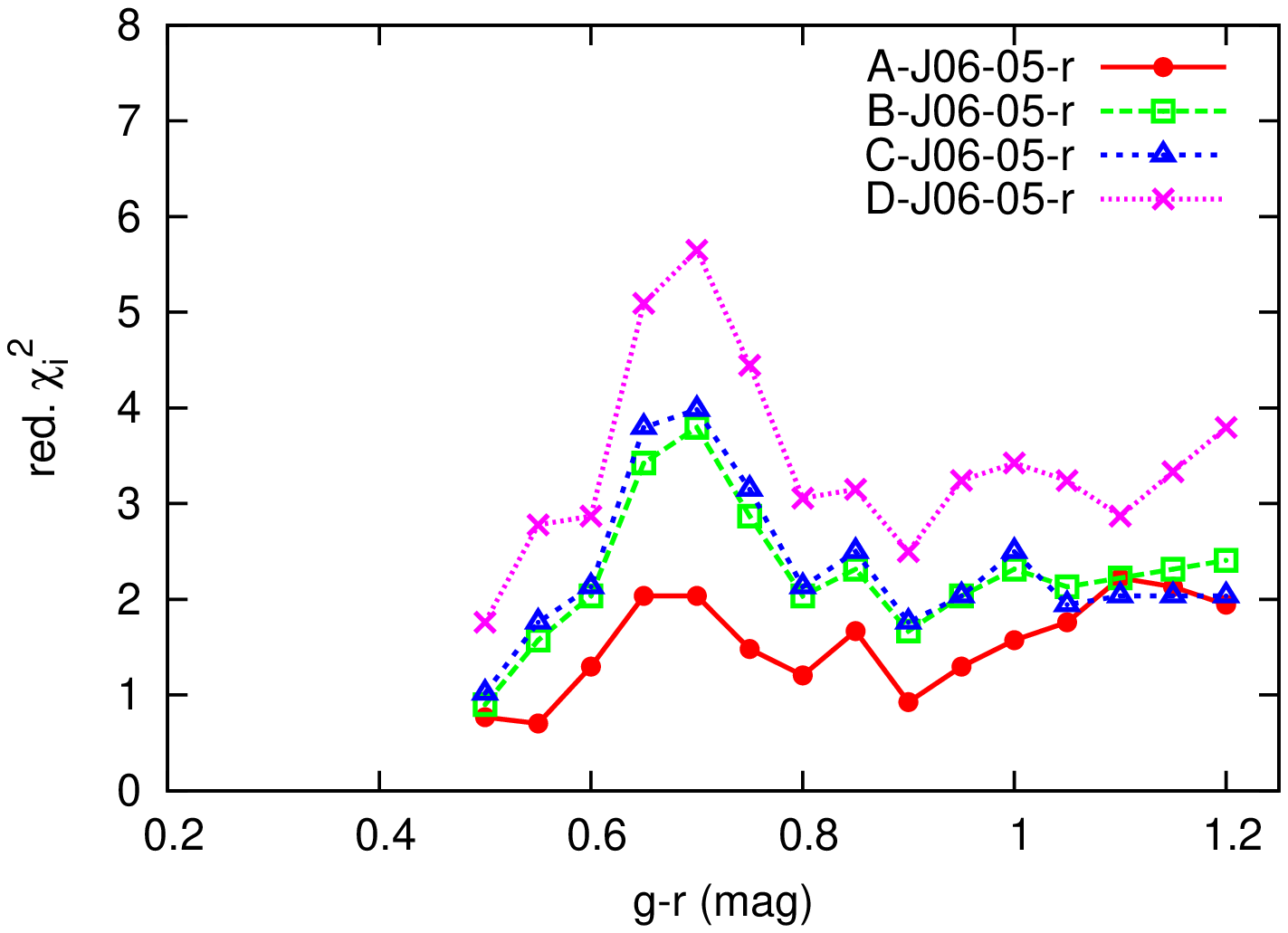}}}
\centerline{\resizebox{0.98\hsize}{!}
{\includegraphics[angle=270]{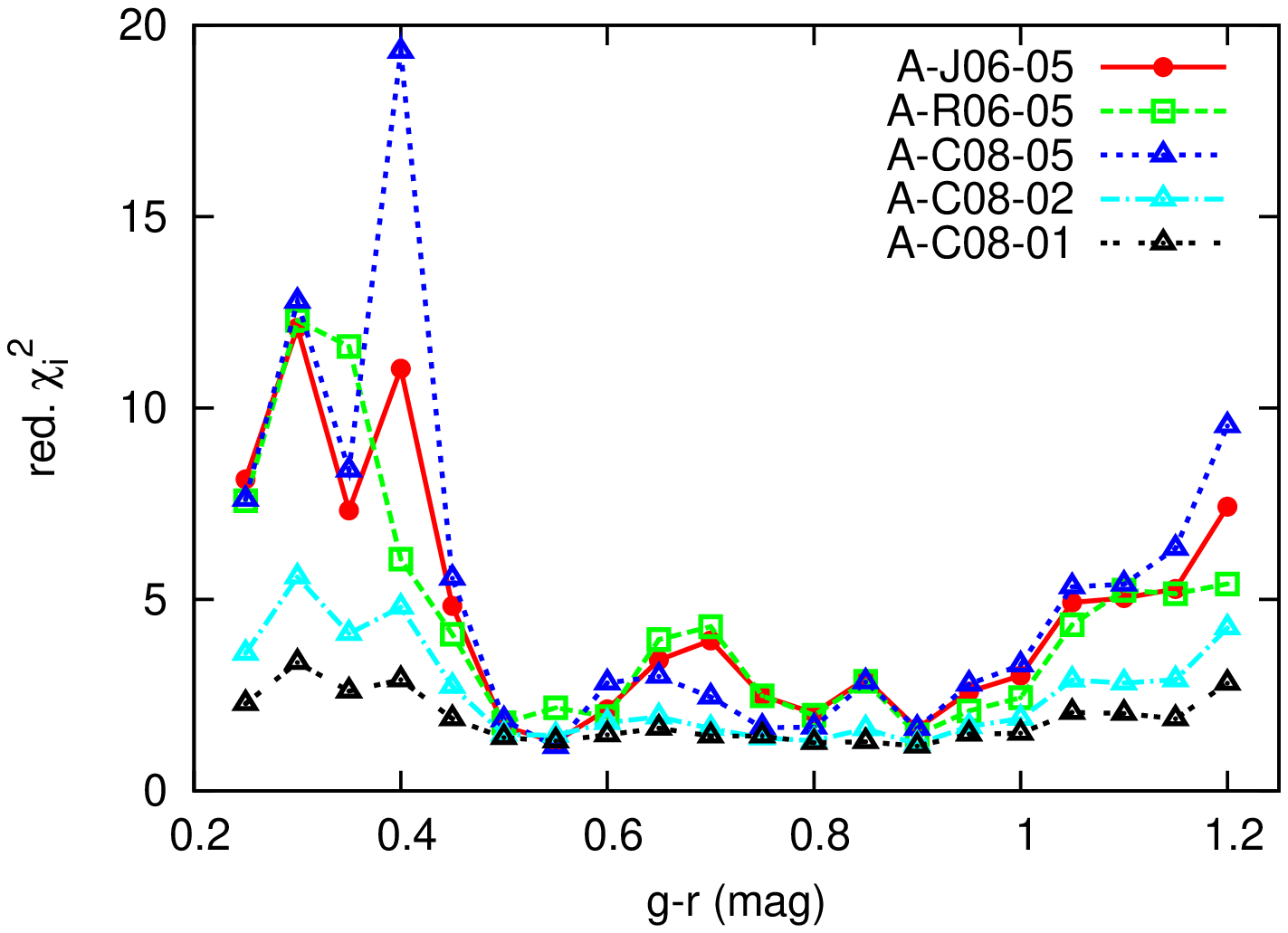}}
}
\caption[]{
Reduced $\chi^2_\mathrm{i}$ values in each colour bin. The top panel shows the comparison
of disc models A -- D. The bottom panel shows the dependence on the filter
transformation and on the smoothing in $g$.
}
\label{figchi2-gr}
\end{figure}

Model A-J06-05-r is the best model with a reduced $\chi^2=2.59$ (see table \ref{tab-fit}).
In model A the deviations of data and model are less than $\pm10$ percent over the
full colour range and in the magnitude regime, where the MS stars of thin disc, 
thick disc or halo dominate. The $\chi^2$ distribution in figure
\ref{figA-D} shows a strong noise
component and the contribution by systematic deviations of
the predicted density profiles from the real profiles. In the top panel of
figure \ref{figchi2-gr} the reduced $\chi^2_\mathrm{i}$ values in each colour bin are
quantified. The contributions to the total $\chi^2$ are fairly uniform.

The main feature of the systematic
discrepancies is the shallow valley in the transition of thin disc
and thick disc in the data, which is not reproduced by the model. 
This is a sign of an even steeper slope of the thin disc above 1\,kpc. 
A continuous transition between the old
thin disc with a maximum vertical velocity dispersion of $\sigma_\mathrm{e}=25$\,km/s and
the thick disc with $\sigma_\mathrm{t}=45.3$\,km/s can be excluded, because it
would add more stars to the transition regime.

In models B, C, and D, the fits in the magnitude range of dominating thin disc
contribution are significantly worse, whereas the thick disc parameters can be
adjusted to reach a similar good fit at fainter magnitudes as in model A. This is also quantified in the
$\chi^2$ values (see table \ref{tab-fit}). Model D with the small disc age and
strongly declining SFR is the worst model with $\chi^2=5.64$. Models B and C are
comparable with $\chi^2=3.75$ and 3.84, respectively.
All three models overestimate
the thin disc density at large heights leading to a deficit closer to the plane.
 A more detailed look on
the $\chi^2_\mathrm{i}$ distributions shows that model A is superior in a wide colour range
(figure \ref{figchi2-gr}).

\begin{figure}
\vspace*{0.5\hsize}
\centerline{\resizebox{0.98\hsize}{!}
{\includegraphics[angle=0]{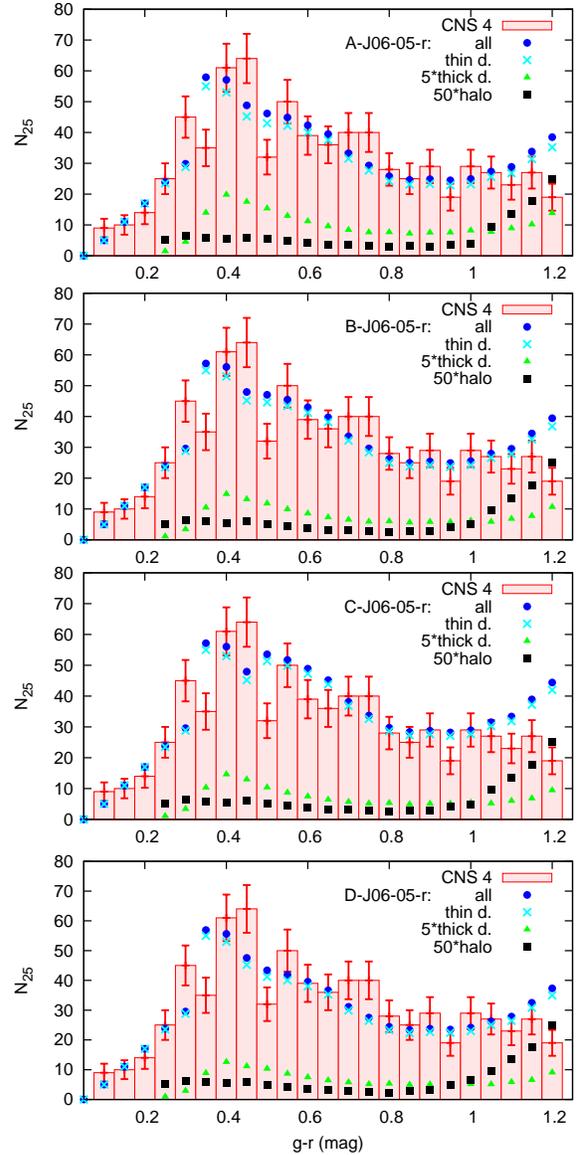}}}
\caption[]{
Local normalisations for disc models A -- D (top to bottom). The histogram shows the data
from the CNS4 and the full blue circles are the best fit values. The
contributions of thin disc, thick disc ($\times 5$), and stellar
halo ($\times 50$) are shown separately.
}
\label{fign-gr}
\end{figure}

Since we used the local number densities of all components as free fitting 
parameters, a comparison of the fitting results and the
observed star counts in the solar neighbourhood is a crucial test of the models. In figure \ref{fign-gr}
the histogram with error-bars shows the data from the CNS4. The full (blue)
circles are the results of model A -- D (from top to bottom). In all cases there is a
reasonable match and also the total number of MS stars agree (see table
\ref{tab-fit}). The observed number of MS stars range from
 $N_\mathrm{25}=726$ (with R06) to $N_\mathrm{25}=770$ (with C08).
 Additionally the contribution of thin disc, thick disc and halo is
plotted separately in figure \ref{fign-gr} (enhanced by some factor to make it visible) demonstrating that
the local number densities of all components are smooth functions of $(g-r)$. 
A more detailed discussion is given in section \ref{sub-filter}.

As a bottom line we find that the local disc model A is fully consistent with the
NGP star count data of SDSS. Model A with a relatively large fraction of stars
older than 8\,Gyr and a correspondingly small maximum vertical velocity 
dispersion of $\sigma_\mathrm{e}=25$\,km/s is the best fitting model. In the next
sub-sections we test the robustness of the result with respect to the filter
transformations used for the solar neighbourhood and with respect to the smoothing
in $g$.

\subsection{Filter transformations}\label{sub-filter}

\begin{figure}
\centerline{\resizebox{0.98\hsize}{!}
{\includegraphics[angle=270]{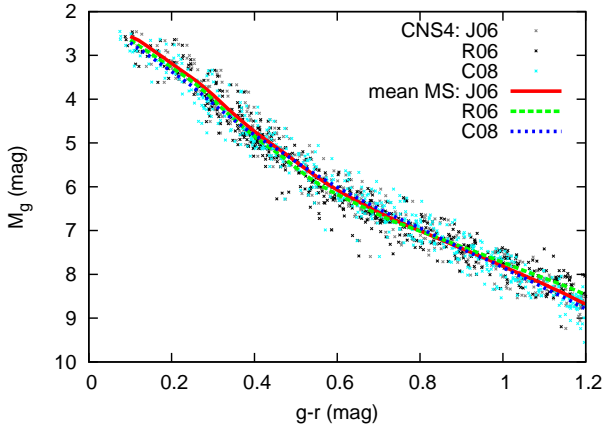}}}
\caption[]{
Local CNS4 stars and mean MS using different filter transformations. Each star
is plotted three times in different colours to show the systematic shift of the
MS.
}
\label{figmsJRC}
\end{figure}

\begin{figure}
\centerline{\resizebox{0.98\hsize}{!}
{\includegraphics[angle=270]{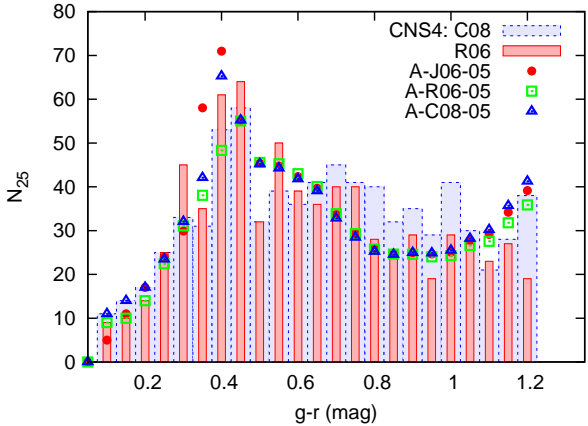}}}
\caption[]{
Local normalisations for disc model A using different filter transformations.
}
\label{fignJRC}
\end{figure}

In figure \ref{figmsJRC} the local MS is shown for three different transformations
(J06 for \citet{Jo06}, R06 for \citet{Ro06}, C08 for \citet{Ch08}). There are small
systematic deviations in $M_\mathrm{g}(g-r)$ which may influence the best fit
parameters of the model. The histograms in figure \ref{fignJRC} show the local
star counts for R06 and C08.
Additionally to the differences in the absolute magnitudes for thin and thick
disc the fraction of thin
disc turn-off stars is reduced in C08 and is set to zero in R06
in order to test their influence on the $\chi^2$ values and number counts.

We present the comparison using the different local MS
for model A. The effects in models B, C and D are very similar. 
First we derived
the best fit models for the reduced colour-magnitude regime as before in
A-J06-05-r for models A-R06-05-r and A-C08-05-r (see table \ref{tab-fit}). 
The total $\chi^2$ values and local normalisations of the three models are very similar. 

\begin{figure}
\centerline{\resizebox{0.9\hsize}{!}{\includegraphics[angle=0]
{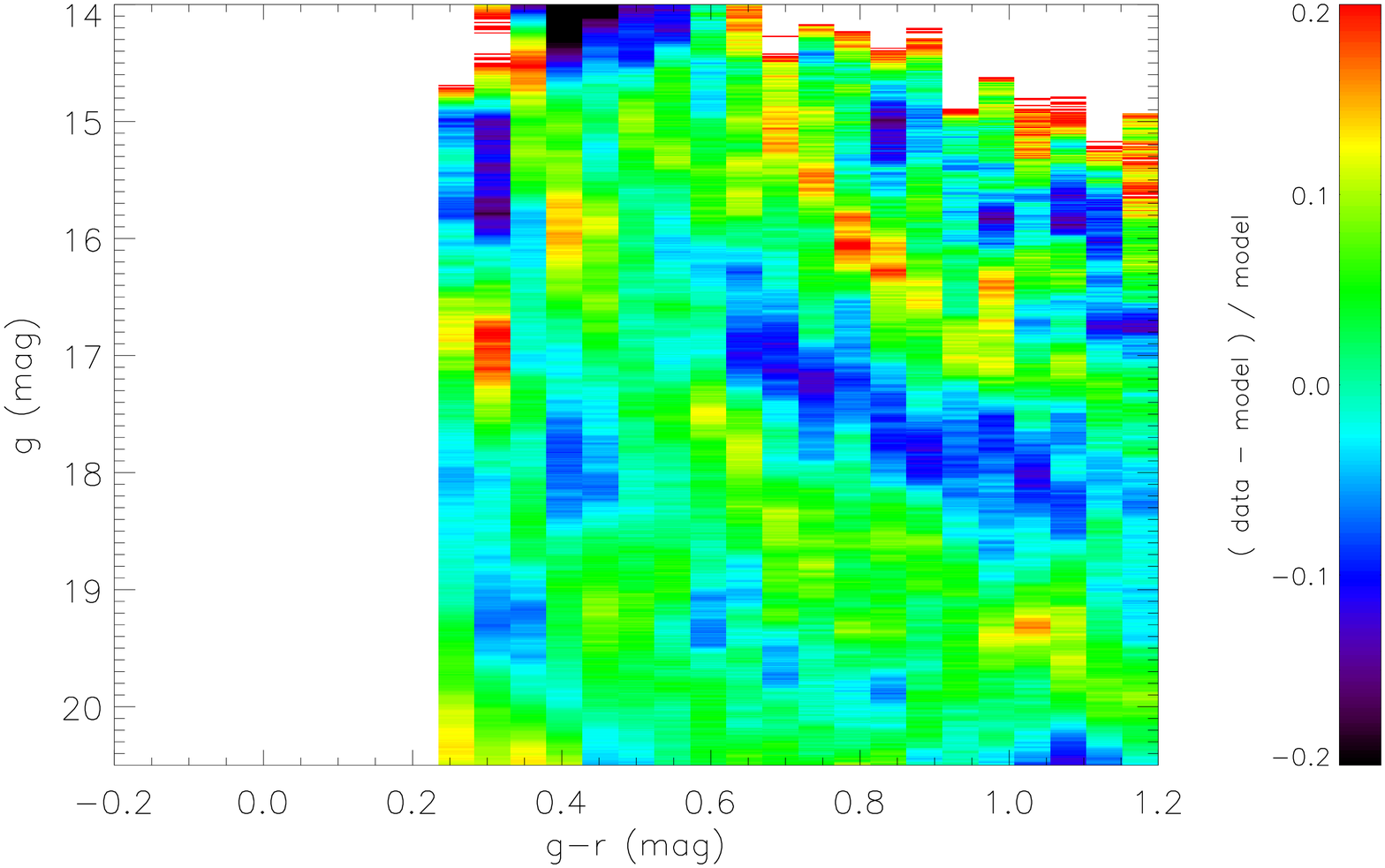}}}
\centerline{\resizebox{0.9\hsize}{!}{\includegraphics[angle=0]
{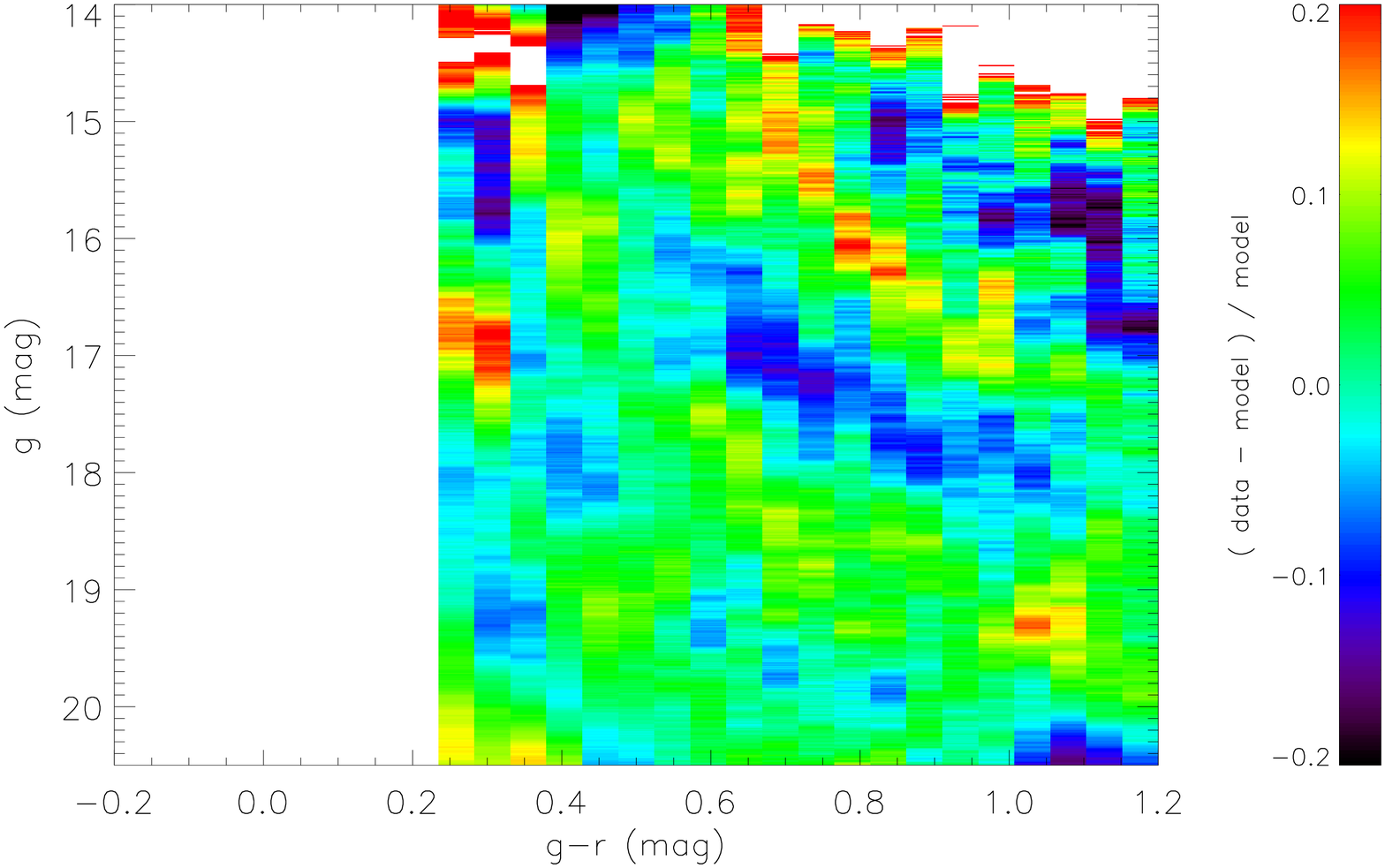}}}
\centerline{\resizebox{0.9\hsize}{!}{\includegraphics[angle=0]
{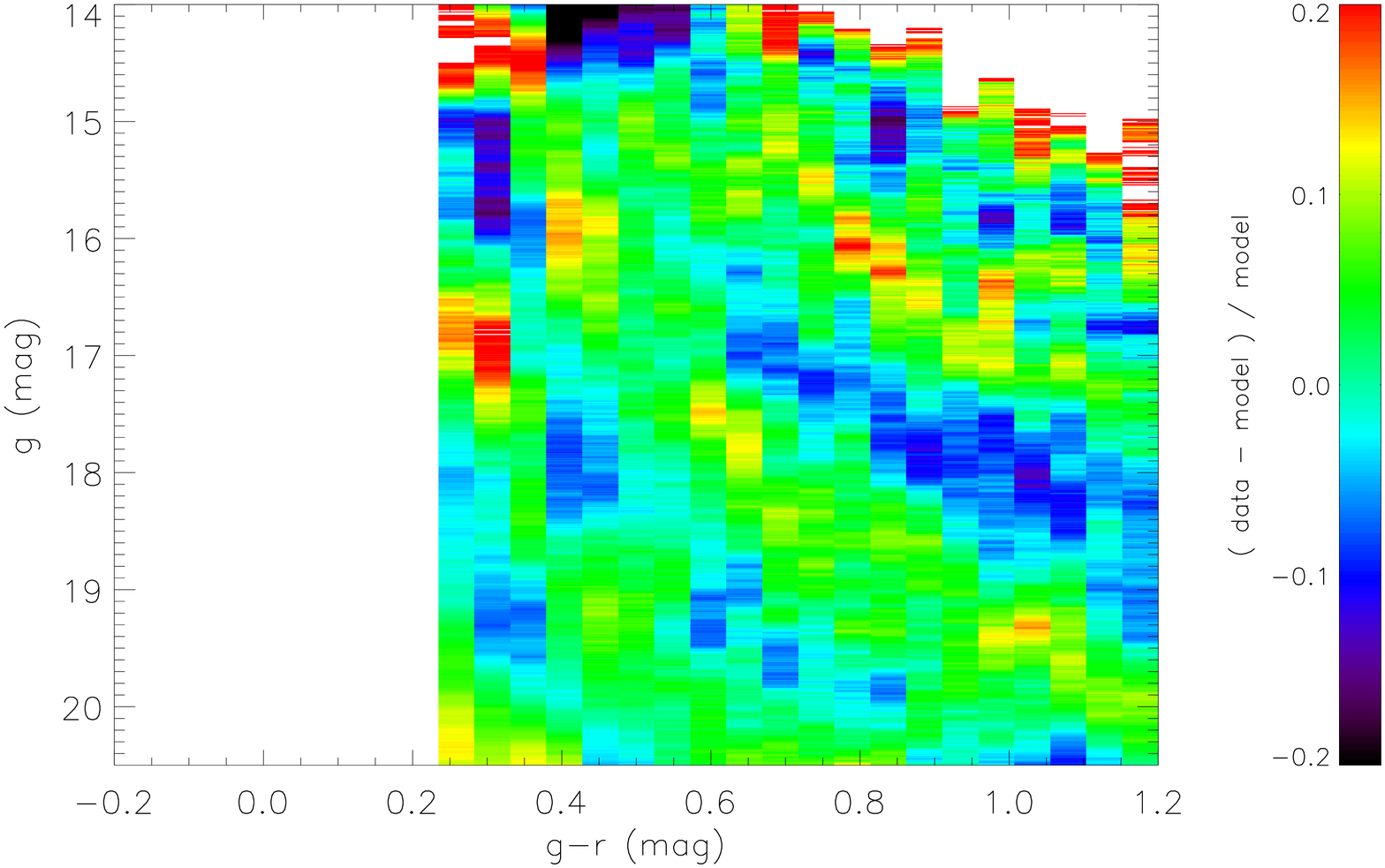}}}
\caption[]{
Relative differences of NGP data and models A-J06-05, A-R06-05, and A-C08-05
with different filter transformations (top to bottom).
}
\label{figdiffJRC}
\end{figure}

Then we extended the
fit regime to the full CMD and recalculated the local normalisations and the
$\chi^2$ values.
Figure \ref{figdiffJRC} shows the relative deviations of data and models.
The patterns of the deviations differ slightly in the regime, where the thin 
disc dominates. 
 The corresponding $\chi^2_\mathrm{i}$ values in the colour bins are shown in
the lower panel of figure \ref{figchi2-gr}. For the extended fit regime there
are small differences arising mainly from the turn-off stars. In the colour bin
$g-r=0.4$ the fraction of turn-off stars is 80 percent (J06), 0 percent (R06)
and 60 percent (C08), respectively. Model A-R06-05 without turn-off stars gives the best
result.  A consequence is that a larger fraction of stars are allocated to the
thick disc instead of the thin disc in model A-R06-05 (see figure
\ref{fignJRC}).
 At $g-r=0.35$ model A-J06-05 with 55 percent turn-off stars gives the
best result and for $g-r\le 0.3$ the turn-off stars are too bright to contribute
to the fit.
At the red end $\chi^2_\mathrm{i}$ is influenced by the missing
giants. Therefore it is not useful to discuss here the differences of the fits in
more detail.

The local normalisations do not differ significantly (see figure \ref{fignJRC}).
The differences in the histograms for the transformations R06 and C08 show the
possible variations in the local star counts due to noise and the effect of a
different slope in the MS leading to a shift of stars in colour. All
transformations are consistent with the local star count data. Only in the
transition of K to M dwarfs the large number of thin disc stars in $g-r=1.2$ may
be a hint that the MS turning point is relatively blue as in C08. 

\subsection{Smoothing}

\begin{figure*}
\centerline{\resizebox{0.3\hsize}{!}{\includegraphics[angle=0]
{g05_gr005_bp900_l1800_db100_dl1800}}
\resizebox{0.3\hsize}{!}{\includegraphics[angle=0]
{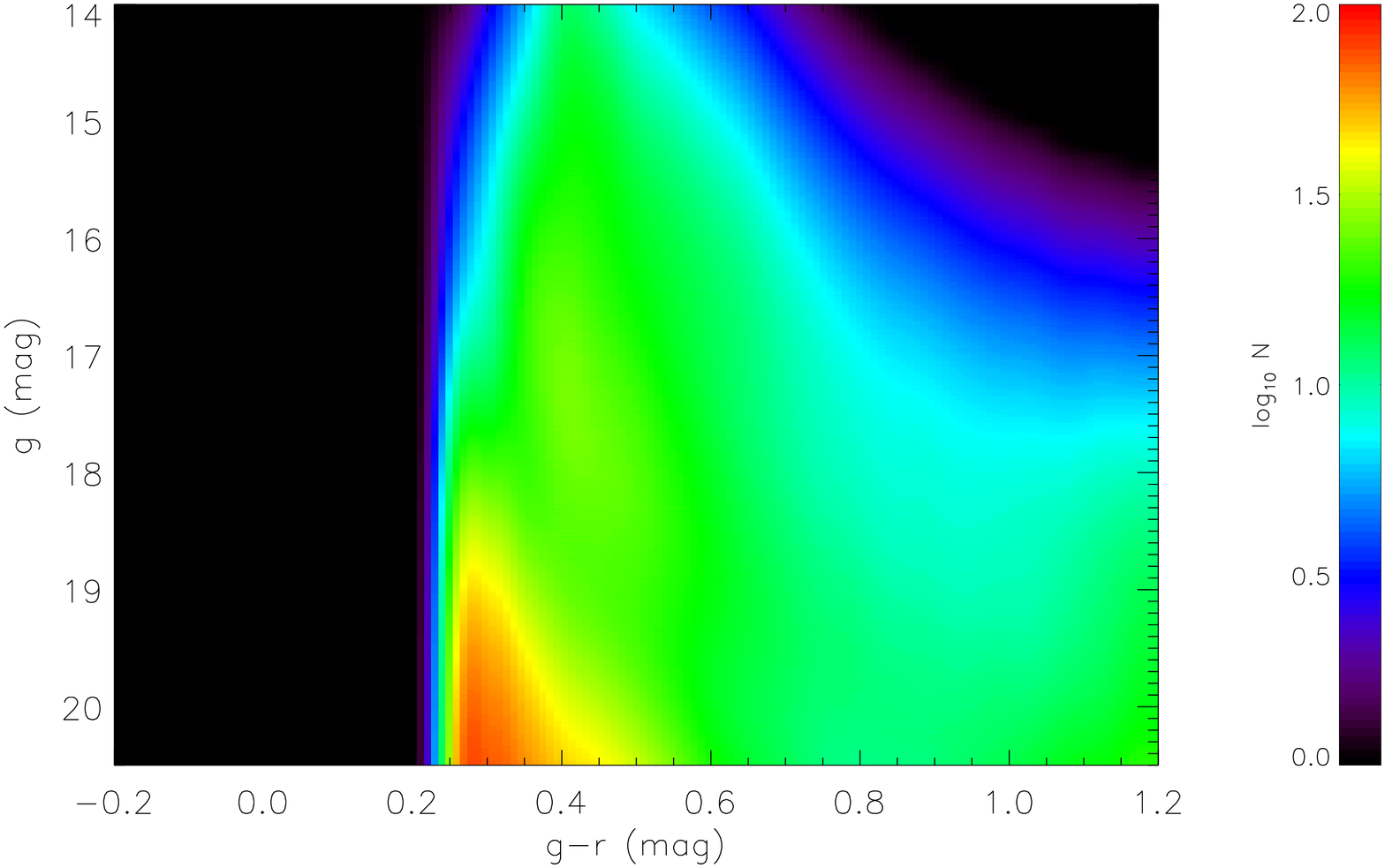}}
\resizebox{0.3\hsize}{!}{\includegraphics[angle=0]
{s9_ft00_mataha07_g05_gr005_bp900_l1800_diff}}}
\centerline{\resizebox{0.3\hsize}{!}{\includegraphics[angle=0]
{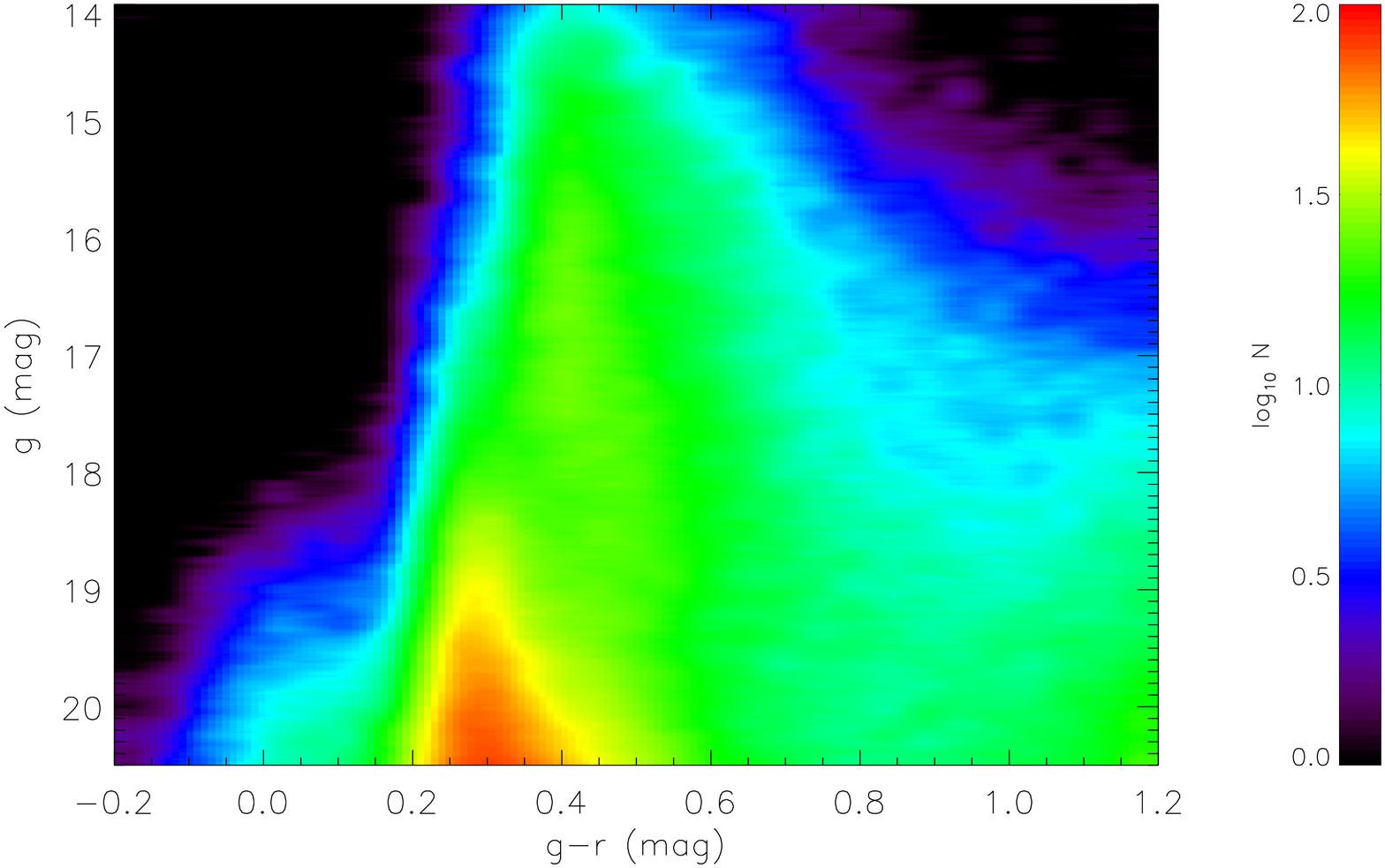}}
\resizebox{0.3\hsize}{!}{\includegraphics[angle=0]
{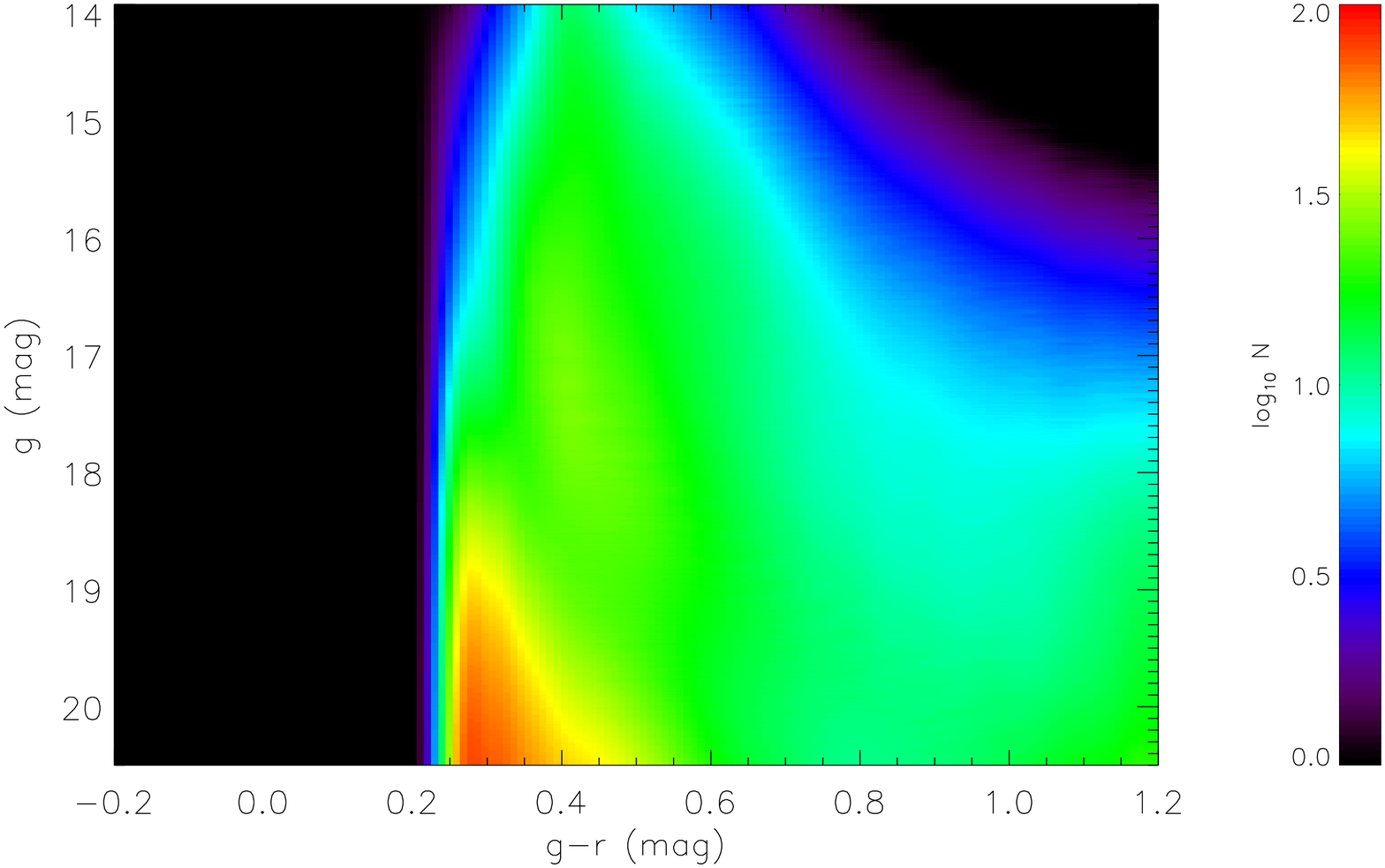}}
\resizebox{0.3\hsize}{!}{\includegraphics[angle=0]
{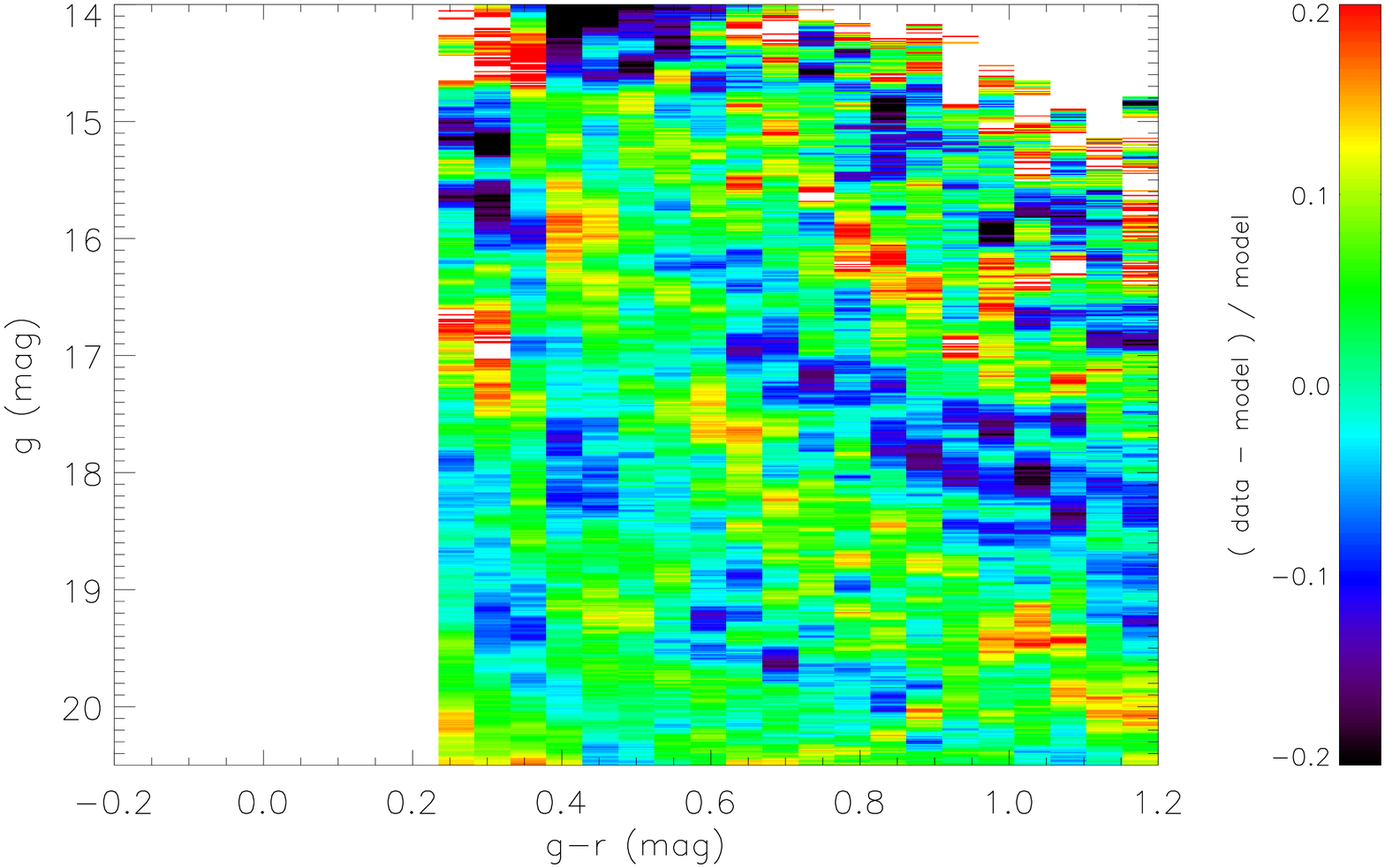}}}
\centerline{\resizebox{0.3\hsize}{!}{\includegraphics[angle=0]
{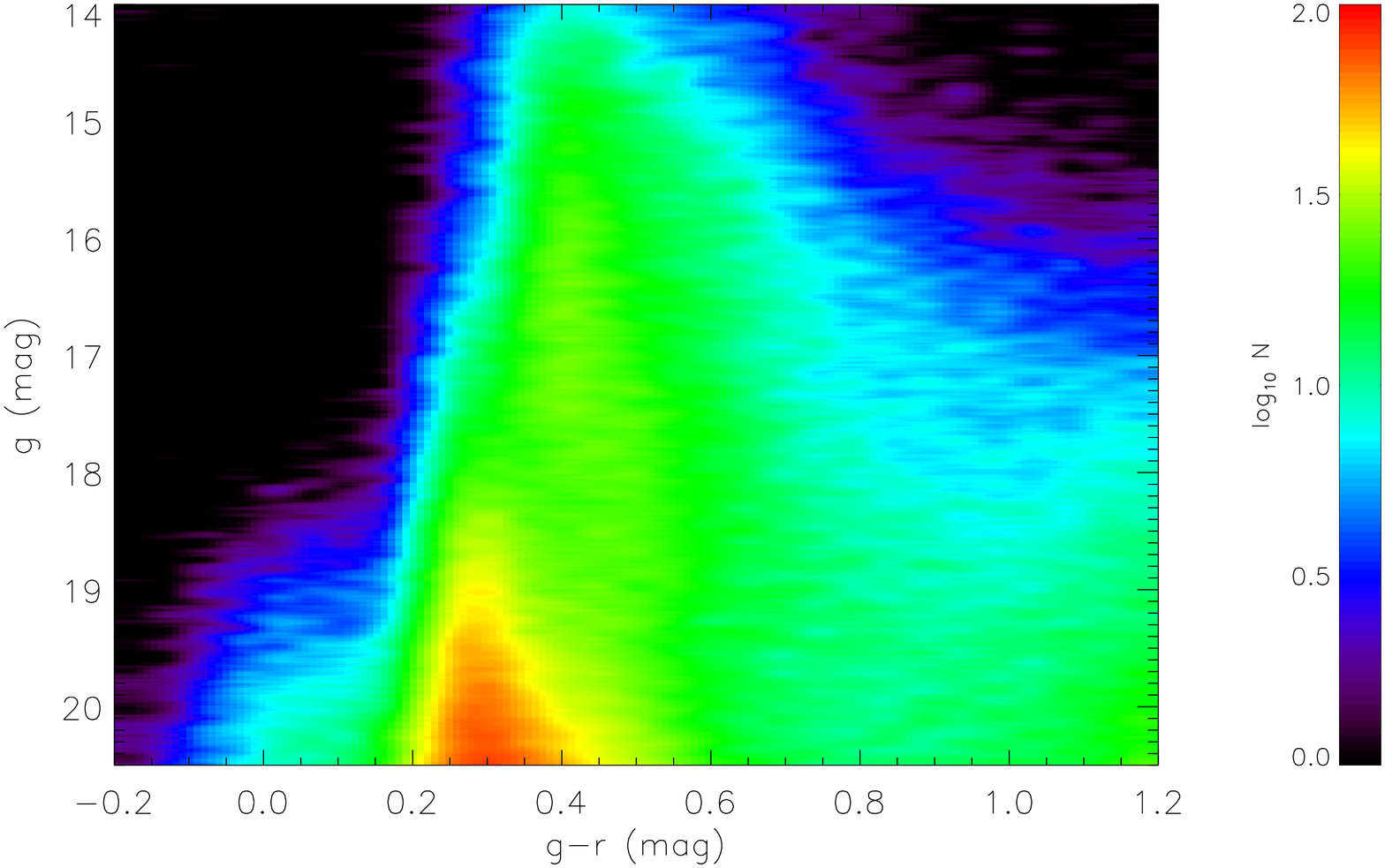}}
\resizebox{0.3\hsize}{!}{\includegraphics[angle=0]
{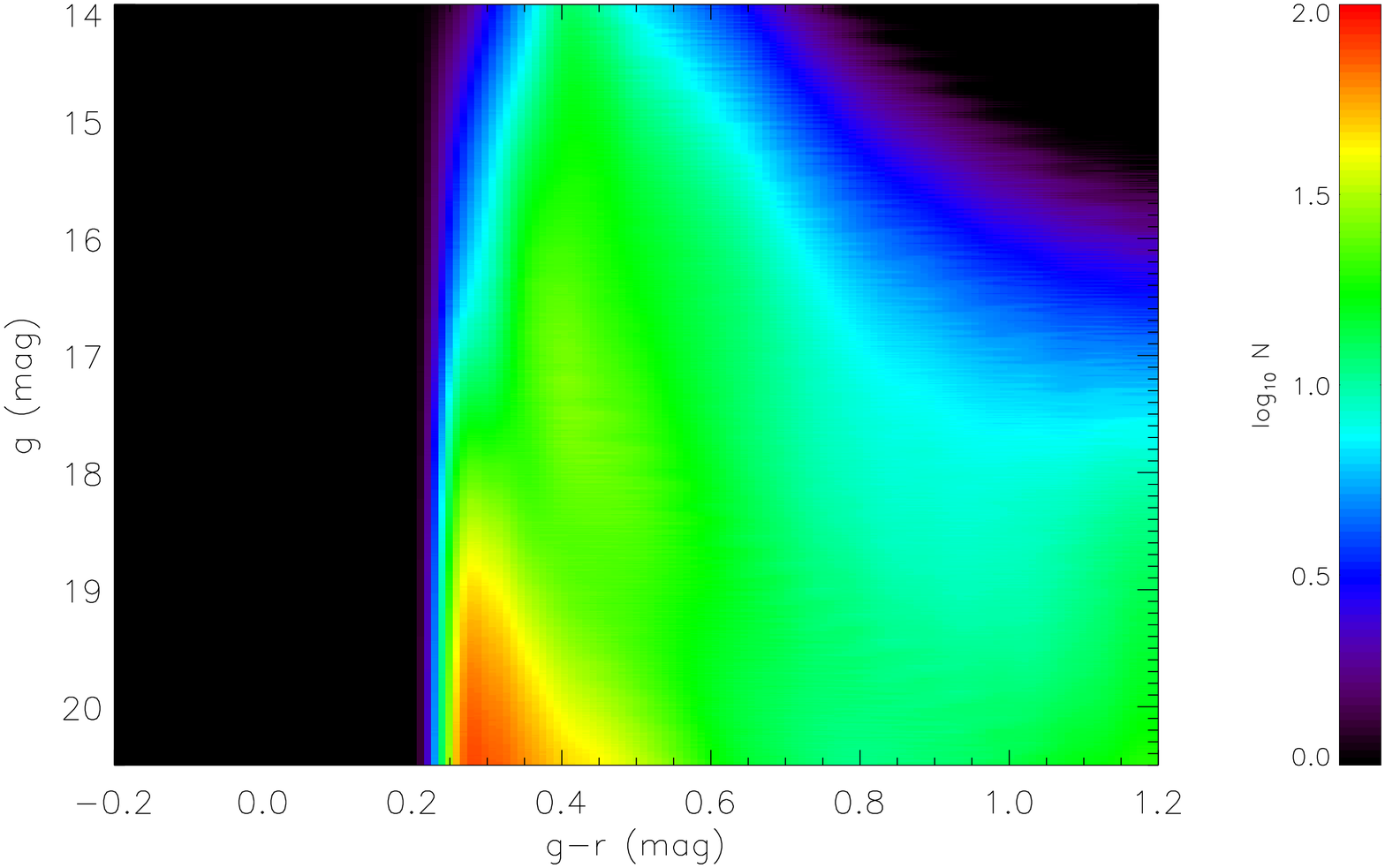}}
\resizebox{0.3\hsize}{!}{\includegraphics[angle=0]
{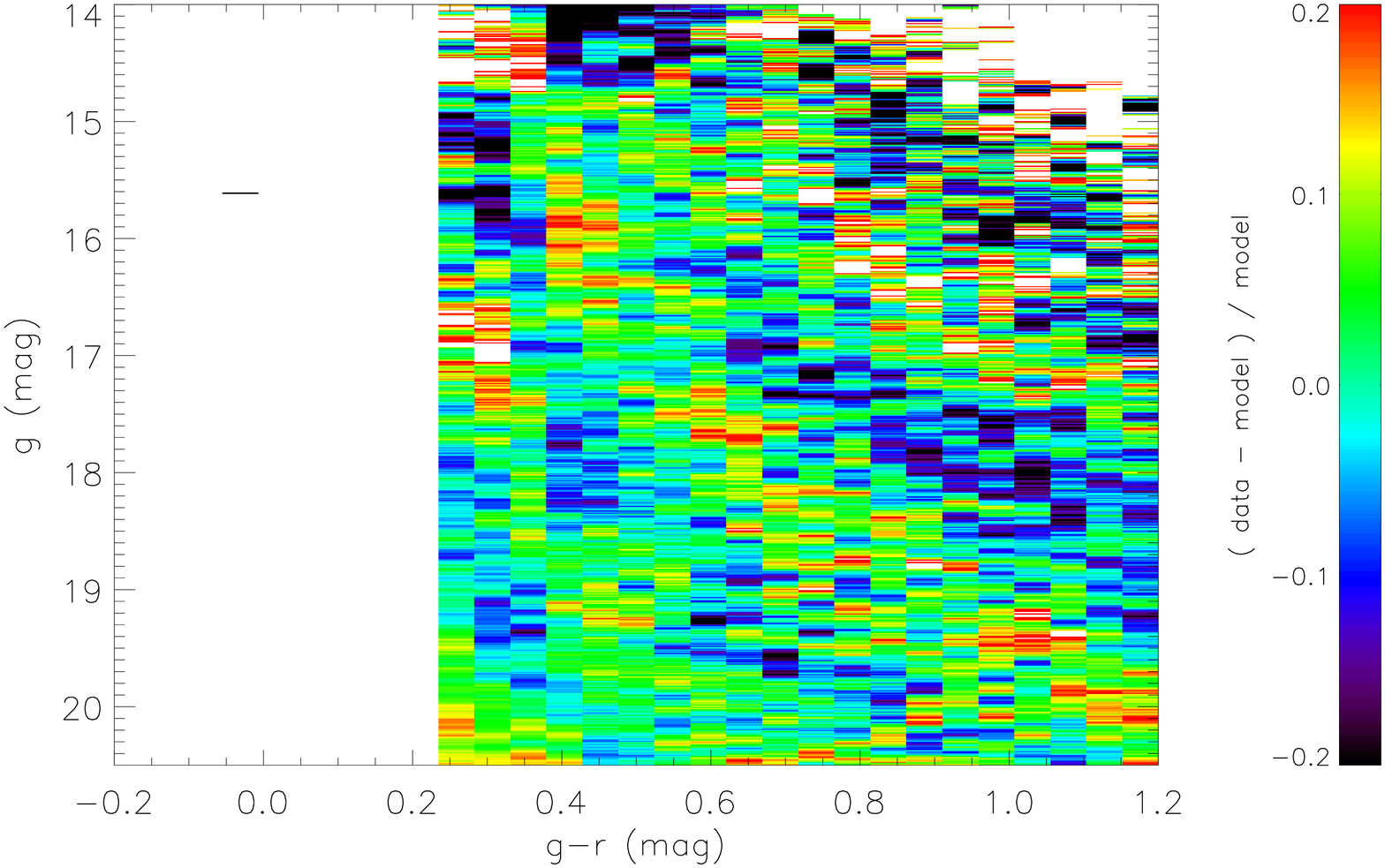}}}
\caption[]{
Decreasing smoothing $\Delta g=$0.5, 0.2, 0.1 (top to bottom).
NGP data (left), models A-C08-05,02,01 (middle), 
relative differences (right). Same
colour coding as in figure \ref{figA-D}.
}
\label{figsmooth}
\end{figure*}

Smoothing of the data has two main effects. On one hand statistical noise
 is reduced. On the other hand physical features
in the data are smeared out and may be shifted systematically.
Figure \ref{figsmooth} shows the NGP data and corresponding models 
A-C08-05,02,01 for different smoothing lengths $\Delta g=$0.5, 0.2, 0.1\,mag, respectively.
The $\chi^2$ values are decreasing due to to the increasing degrees of freedom
(see table \ref{tab-fit} and figure \ref{figchi2-gr}). The local normalisations
are statistically indistinguishable. Only in the colour bin $g-r=0.4$ a few
percent of thin disc stars are shifted to the thick disc at higher
resolution reducing the contribution to $\chi^2_\mathrm{i}$ by the
bright magnitude bins dramatically.

\section{Conclusions}\label{sec-conclusion}

We have used four different models with different SFRs of the thin disc, which 
fit the local kinematics of main sequence (MS) stars (Paper I) and compared the star 
count predictions for the North Galactic Pole (NGP) field with $b>80^0$ of the Sloan
Digital Sky Survey (SDSS). For the
thin disc we applied the absolute magnitudes of the local MS as determined in 
\citet{Ju08}.  A self-consistent isothermal thick disc and a simple stellar halo model
was added to complete the contributions of MS stars to the star counts.
We used the local normalisations of thin disc, thick disc and halo in each 
colour bin as
free parameters and minimise $\chi^2$ in the Hess diagram over a large
colour-magnitude range in $(g-r,g)$.

All models match the observed number densities better than $\pm 20$ percent
proving the reliability of the thin disc density profiles 
by the self-consistent disc models in the range of $|z|<1$\,kpc. The derived
local normalisations are consistent with the star count data of the CNS4 in the
solar neighbourhood.
The $\chi^2$ analysis shows that model A is clearly preferred with systematic
deviations of a few percent only. The SFR of model A is characterised by a
maximum at an age of 10\,Gyr and a decline by a factor of four to the present day
value of 1.4\,$\msun$/pc$^2$/Gyr. In the thin disc the present day fraction of
stars older than 8\,Gyr is with 54 percent significantly higher than for models
B, C, and D (Paper I). Especially model C with a constant SFR and model D
with a disc age of 10\, Gyr can be ruled out.

The thick disc can be modelled very well by an isothermal simple stellar
population. The density profile can be approximated by a
sech$^{\alpha_\mathrm{t}}(z/\alpha_\mathrm{t}h_\mathrm{t})$ function. For model
A we find a power law index of $\alpha_\mathrm{t}=1.16$ in-between an exponential
profile and a sech$^2$-profile (the latter corresponds to an isolated isothermal disc). 
The exponential scale height
is $h_\mathrm{t}=800$\,pc corresponding to a vertical velocity dispersion of 
$\sigma_\mathrm{t}=45.3$\,km/s. About 6 percent of the stars in the solar
neighbourhood are thick disc stars. In \citet{Jur08} the stellar density 
distribution in the Milky Way based on SDSS star counts was fitted by 
exponential thin and thick disc profiles. The result is a much larger thick disc
scale height, which balances the flattening of the profile at low $z$.

The results do not depend significantly on the filter transformations used for
the local MS nor on the smoothing of the data in luminosity. For the future an
extension of the model to include turn-off stars in more detail and the
contribution of giants as well as a higher resolution in colour at the blue end
would be very useful.

We also plan to apply the model to lower Galactic latitudes in order to determine radial scale lengths of thin and thick disc as well as radial gradients in the stellar populations.

\section*{Acknowledgements}

SG is supported by a grant of the Chinese Academy of Science.

SV was supported by an Alexander von Humboldt Fellowship
(Web Site http://www.humboldt-foundation.de).

Funding for the SDSS and SDSS-II has been provided by the Alfred P. Sloan 
Foundation, the Participating Institutions, the National Science Foundation, 
the U.S. Department of Energy, the National Aeronautics and Space 
Administration, the Japanese Monbukagakusho, the Max Planck Society, and the 
Higher Education Funding Council for England. The SDSS Web Site is 
http://www.sdss.org/.

The SDSS is managed by the Astrophysical Research Consortium for the 
Participating Institutions. The Participating Institutions are the American 
Museum of Natural History, Astrophysical Institute Potsdam, University of 
Basel, University of Cambridge, Case Western Reserve University, University of 
Chicago, Drexel University, Fermilab, the Institute for Advanced Study, the 
Japan Participation Group, Johns Hopkins University, the Joint Institute for 
Nuclear Astrophysics, the Kavli Institute for Particle Astrophysics and 
Cosmology, the Korean Scientist Group, the Chinese Academy of Sciences 
(LAMOST), Los Alamos National Laboratory, the Max-Planck-Institute for 
Astronomy (MPIA), the Max-Planck-Institute for Astrophysics (MPA), New Mexico 
State University, Ohio State University, University of Pittsburgh, University 
of Portsmouth, Princeton University, the United States Naval Observatory, and 
the University of Washington. 

\end{document}